\documentclass[revtex4]{emulateapj}

\usepackage{graphicx}
\usepackage{lscape}
\usepackage{amssymb}
\usepackage{epstopdf}
\usepackage[flushleft]{threeparttable}
\usepackage{amsmath}
\usepackage{mathrsfs}
\usepackage{float}
\usepackage{natbib}
\usepackage{longtable}
\usepackage{rotating}
\usepackage{footmisc}

\begin{document}

\shorttitle{Abundances of Metal-Poor Stars}
\shortauthors{O'Malley, McWilliam, Chaboyer, Thompson}

\title{A Differential Abundance Analysis of Very Metal-Poor Stars \footnote{This paper includes data gathered with the 6.5-m Magellan Clay Telescope located at Las Campanas Observatory, Chile.} \footnote{Some of the data presented herein were obtained at the W.M. Keck Observatory, which is operated as a scientific partnership among the California Institute of Technology, the University of California and the National Aeronautics and Space Administration. The Observatory was made possible by the generous financial support of the W.M. Keck Foundation.}}

\author{Erin M. O'Malley}
\affil{Department of Physics and Astronomy, Dartmouth College, Hanover, NH 03784}
\email{Erin.M.O'Malley.GR@dartmouth.edu}

\author{Andrew McWilliam}
\affil{Observatories of the Carnegie Institution of Washington, Pasadena, CA 91101}

\author{Brian Chaboyer}
\affil{Department of Physics and Astronomy, Dartmouth College, Hanover, NH 03784}

\author{Ian Thompson}
\affil{Observatories of the Carnegie Institution of Washington, Pasadena, CA 91101}

\begin{abstract}
We have performed a differential, line-by-line, chemical abundance analysis, ultimately relative to the Sun, of nine very metal-poor main sequence halo stars, near [Fe/H]=$-$2\,dex.  Our abundances range from $-2.66\leq\mathrm{[Fe/H]}\leq-1.40$\,dex with conservative uncertainties of 0.07\,dex.  We find an average [$\alpha$/Fe]$=0.34\pm0.09$\,dex, typical of the Milky Way.  While our spectroscopic atmosphere parameters provide good agreement with HST parallaxes, there is significant disagreement with temperature and gravity parameters indicated by observed colors and theoretical isochrones.  Although a systematic underestimate of the stellar temperature by a few hundred degrees could explain this difference, it is not supported by current effective temperature studies and would create large uncertainties in the abundance determinations.  Both 1D and $\langle$3D$\rangle$ hydrodynamical models combined with separate 1D non-LTE effects do not yet account for the atmospheres of real metal-poor MS stars, but a fully 3D non-LTE treatment may be able to explain the ionization imbalance found in this work.
\end{abstract}

\keywords{Very metal-poor stars; Abundances; Parallaxes}

\section{Introduction}
One of the major efforts in current research is to determine a more accurate age for the Milky Way as an independent test of modern theories of cosmology.  Metal-poor globular clusters (GCs) are important in this effort as they are among the oldest objects in the Galaxy.  A comparison of GC color-magnitude diagrams (CMDs) to theoretical isochrones can be used to determine GC distances and  ages.  Stellar evolution models were first developed for solar-metallicity stars but are now pushing towards calibration at lower metallicities.  The calibration of these models requires one to define the location of the main-sequence (MS) as a function of metallicity via MS-fitting; however, to date there is only one star, HD\,25329, with [Fe/H] $<$ -1.4 dex and with an \emph{Hipparcos} parallax \citep{Leeu2007} and chemical abundance determination from high resolution spectroscopy \citep{BS1994} accurate enough for use in MS-fitting.  This lack of calibration stars introduces a great amount of uncertainty in the distance and age determinations for metal-poor GCs which we we aim to improve with our study.

One of the key features for an accurate MS-fitting calibration is accurate distances and abundances for very low metallicity MS stars.  Accurate parallaxes for metal-poor stars are now available in the first \emph{Gaia} data release, which use the Tycho-Gaia astrometric solution (TGAS) \citep{MLH2015,Gaia,Lindegren2016} to obtain parallaxes with typical uncertainties of 0.25 mas; however, even more accurate stellar parallaxes have been measured with the Hubble Space Telescope's (HST) fine guidance sensor (FGS) for nine very metal-poor dwarf stars ($\sigma_\pi = 0.11$) and are presented in \citet{Chab2016}.  Our goal is to obtain accurate overall metallicities of the same very metal-poor MS stars with HST parallaxes to be used in MS-fitting. 

We were motivated to use the robust technique of a line-by-line, differential abundance analysis, ultimately relative to the Sun, as it should provide abundances that are independent of log\,gf values.  The iterative nature of this method was expected to provide excitation temperatures independent of line equivalent widths (EWs) and ionization balance for ionized and neutral species of the same elements.  However, in the process we uncovered issues (e.g. non-ionization equilibrium and low spectroscopic gravities) that a 1D LTE analysis is incapable of resolving.

This paper is organized as follows.  In $\S$2 we present our target selection criteria and spectroscopic observations, in $\S$3 we present the abundance analysis based on the EWs of Fe and $\alpha$-element spectral lines and stellar model atmospheres.  The differential abundance results are provided in $\S$4 along with an error analysis and a comparison to the Dartmouth Stellar Evolution Program isochrones \citep[hereafter DSEP]{Dott2008}.  Finally, the significance of our findings is discussed in $\S$5.

\section{Observations \label{sec:Obs}}
\subsection{Target Selection}
The release of the \emph{Hipparcos} catalog \citep{hip} led to a renewed interest in using MS-fitting to determine distances to GCs \citep[e.g.][]{reid97,gratton97,pont98,chaboyer98,Carr2000,grundahl02,Grat2003}.  These studies generally found that the GCs were somewhat more distant, and hence younger than found in previous work.  The location of the MS in a color-magnitude diagram is a sensitive function of the composition of the stars, and these main sequence fitting studies were hampered by the lack of very metal-poor stars  ($\mathrm{[Fe/H]}\la-1.4$\,dex) which had accurate \emph{Hipparcos} parallaxes.

The original \emph{Hipparocs} parallaxes were reanalyzed by \citet{Leeu2007} leading to smaller uncertainties in the parallaxes.  Unfortunately, even with this improved data, only one very metal-poor star (HD\,25329) was found to have a parallax measured to better than 10\% accuracy, suitable for MS-fitting.  Having only one comparison star at low metallicity results in large uncertainties in distance determinations and therefore ages of metal-poor GCs.  To alleviate this situation, we were awarded time on HST to obtain FGS parallaxes of nine very metal-poor stars with the goal of extending the range of metallicities below at least [Fe/H] $=-2.3\,$dex for stars with well-determined parallaxes.

To select stars for the HST program,  single, un-evolved (i.e. main sequence) stars that were very metal-poor were selected from the lists complied by \citet{Carr2000}, \citet{Grat2003} and \citet{Lath2002}. Based upon the increased number of stars with reliable parallaxes, and the fact that the HST parallaxes would have substantially smaller uncertainties than the \emph{Hipparcos} parallaxes, we anticipated that the HST parallaxes would result in an improvement in the distance scale to metal-poor GCs by a factor of three compared to previous work.  Long term radial velocity monitoring from \citet{Lath2002} was used to ensure the target stars are not members of multiple star systems. To ensure that the stars are on the main-sequence, we required that the target stars  be faint ($M_V>5.5$) and/or have a large surface gravity ($\mathrm{log\,g}>4.4$ dex~cm~s$^{-2}$ (henceforth, dex)) and also be fairly red in color ($\mathrm{B-V} < 0.55$), all characteristics of MS stars.   These specific cut-offs in absolute magnitude, surface gravity and color were based upon an examination of the properties of metal-poor MS stars in the DSEP theoretical isochrones.  Since the \emph{Hipparcos} parallaxes have relatively large uncertainties for these stars, surface gravity and color constraints helped to eliminate evolved stars (sub-giants and giant branch stars) from the sample.

\subsection{Spectroscopic Observations}
High resolution spectroscopy of the nine target stars were obtained between 2008 and 2012 using the Magellan Inamori Kyocera Echelle (MIKE) double spectrograph on the 6.5 meter Magellan II Clay telescope, at Las Campanas Observatory, and the High Resolution Echelle Spectrometer (HiRES) on the twin telescopes at the Keck Observatory.  

Our MIKE spectra roughly cover the range of 3375--5000\AA\ on the blue side and 4850--9160\AA\ on the red side; however, small changes in the choice of grating tilt between runs can lead to the inclusion or exclusion of an order.  The MIKE blue-channel spectra actually extend beyond 5000\AA\ , but beyond 5000\AA\ the spectra are not useful for abundance analysis, due to the presence of spectral ghosts; these ghosts result from internal reflections within the dichroic optical element. No spectral ghosts are detectable in the red-channel spectra.  For all MIKE observations we employed a 0.5$\times$5.0 arc second slit, with 2$\times$1 binning, resulting in a measured spectral resolving power of $R$=48,000 for the red side and $R$=55,000 for the blue side.

The HiRES spectra consist of data from three CCDs in the plane of this single spectrograph; for our setup chips~1 through 3 cover the spectral range 4115--5600\AA\ , 5650--7175\AA\ , and 7235--8695\AA\, respectively.  Our HiRES observations employed a slit width of 0.574 arc seconds and 2$\times$1 binning of the CCD pixels, which resulted in a measured average resolving power of R=70,500; however, significant degradation of the resolving power was seen from the center to the edges of the spectral orders.  We note that the resolving power we have measured, for both spectrographs, is higher than cited in the manuals; this was likely a consequence of the good seeing at the time of observation.

We measured the per pixel S/N from the rms scatter of the blaze peak near 6730--6750\AA\, using the IRAF {\em splot} routine; typically, our S/N values greatly exceed 100:1, permitting reliable measurement of EWs down to the 5m\AA\ level.  Radial velocities (RVs) of the target stars were determined by cross-correlating our spectra with the Kurucz solar spectrum, using the IRAF {\em fxcor} routine; the measurement uncertainties ranged from 0.6 to 0.9 km/s.  Heliocentric corrections to the measured velocities were obtained using the IRAF {\em rvcorrect} routine.  A log of the spectroscopic observations along with HST F606W magnitudes and parallaxes appears in Table~\ref{table:Obs}. 

\begin{table*}[t]
\centering
\caption{Target Star Observations \label{table:Obs}}
\begin{tabular}{r l r r r r r}
\tableline\tableline
HIP\,ID & Telescope & Date\hbox{\hskip0.8cm} & S/N\tablenotemark{1} & F606W & HST $\pi$ & RV$_{helio}$\\
 & & & (pixel$^{-1}$) & (mag) & (mas) & (km/s)\\
\tableline
 46120 & Magellan  & 25 May 2009 & 123.95 &  9.938  & 15.011 & $-$95.2 \\
 54639 & Magellan  & 25 May 2009 & 145.80 &  11.149 & 11.116 & $+$63.4 \\
 66815 & Magellan  &  1 June 2005 & 254.0 &  ...   & ...    & $-$64.5 \\
 80679 & Keck & 10 Oct 2008 & 78.17 &  ...    &  ...           & $-$308.3 \\
 87062 & Keck &  9 Oct 2008 & 150.54 &  10.379 & 8.205  & $+$84.5 \\
 87788 & Magellan  & 26 May 2009 & 76.50 &  11.109 & 10.830 & $-$205.6\\
 98492 & Keck &  9 Oct 2008 &  133.03 &  11.377 & 3.487  & $-$266.4\\
103269 & Keck & 10 Oct 2008 & 172.48 &  10.084 & 14.118 & $-$130.5\\
106924 & Keck & 11 Oct 2008 & 131.06 &  10.156 & 14.474 & $-$244.5\\
108200 & Magellan  & 14 Aug 2012 & 92.17 &  10.785 & 12.397 & $-$184.2\\
\hline
\end{tabular}
\tablenotetext{1}{S/N measured from rms scatter at blaze peak near 6740\AA.} 
\end{table*}

\section{Abundance Analysis \label{sec:Abund}}
We performed model atmosphere abundance analysis of our program stars in an iterative procedure, where the stellar temperatures (T$_{\rm eff}$), [Fe/H], [$\alpha$/Fe] and microturbulent velocities were determined from the EWs of lines measured from the high S/N spectra.  We used these T$_{\rm eff}$, [Fe/H] and [$\alpha$/Fe]  values, together with published photometric data, to determine the log\,g indicated by the DSEP theoretical isochrones.  A check on the adopted isochrone log\,g is obtained from the [Fe/H] determined from Fe~II lines, which are sensitive to the model atmosphere gravity.

In order to reduce systematic errors on [Fe/H] and [$\alpha$/Fe] due to inaccurate line oscillator strengths (log\,gf values), we employed a line-by-line differential abundance analysis, relative to the same lines in the solar spectrum, similar to the method employed by \citet{KM2008}.  This method may also provide some mitigation for a number of systematic errors, due to incomplete input physics in the analysis (e.g. blends, non-LTE effects, 3D hydrodynamical effects).  This differential analysis is facilitated by the fact that the gravities of the halo dwarfs are close to the solar value, while their temperatures are not too far from the solar T$_{\rm eff}$ of 5777\,K \citep[e.g.][]{AQ2000}; however, irradiance measurements of \citet{KL2011} indicate 5772 K for the Sun.

Although we determined spectroscopic model atmosphere effective temperatures, we also checked for consistency with temperatures indicated by photometric colors and color-T$_{\rm eff}$ relations.  The photometric values also provided initial T$_{\rm eff}$ estimates for the iterative abundance analysis.

Since unsaturated lines, on the linear portion of the curve of growth, have the greatest sensitivity to abundance, an ideal differential abundance analysis would employ only such weak, unsaturated, lines in both the Sun and the metal-poor program stars.  Typically, unsaturated lines in solar-temperature dwarf stars have an equivalent width below $\sim$30--50m\AA.  However, lines that are 30--50m\AA\ in the Sun are in the 1--4m\AA\ range for our metal-poor dwarf stars.  The need to measure the EW of such weak lines to high precision motivated the acquisition of very high S/N spectra for our metal-poor halo dwarf sample.  Ultimately however, the differential abundance scatter obtained was higher than we desired and the number of 1--4m\AA\ lines was insufficient to adequately constrain the atmosphere parameters and final [Fe/H] for our program stars.  

To address this problem, we chose to employ a standard star, with similar T$_{\rm eff}$ and log\,g to the Sun, but intermediate [Fe/H], in order to increase the number of unsaturated lines in the differential analysis.  Fortunately, we had a high S/N spectrum of the nearby metal-poor dwarf, HIP\,66815, suitable for this purpose.  We also elected to include some stronger lines, which required an estimate of the microturbulent velocity.

One should note that the method of spectroscopic determination of stellar atmosphere parameters is not universally accepted.  \citet{Sitnova2015} caution against use of excitation equilibrium for determining stellar surface temperatures due to NLTE effects and instead opt to use infrared flux method temperatures.  Specifically, they find that their differential approach, relative to the Sun, is only capable of removing abundance trends as a function of excitation potential (EP) for stars with [Fe/H]$\geq-1.5$ dex and state that, for more metal-poor stars, the remaining trend is likely due to uncertainties associated with Van der Waals damping.

There are important differences between \citet{Sitnova2015} and this study that are able to address these issues.  \citet{Sitnova2015} choose an upper limit for their EW measurements of 180 m\AA\ as an attempt to minimize the influence of damped lines on their results.  As can be seen by the results of their differential analysis for metal-poor stars, this upper limit is still too high and their results are in fact influenced by the effects of damping.  On the other hand, we use an upper limit of only 85 m\AA\ in our standard star, HIP\,66815, and our metal-poor MS stars, which corresponds to an upper limit of 120 --130 m\AA\ in the Sun.  The authors also only use lines with EP$>2$ eV and see that when they include lines with lower EP they are able to remove the trend in abundance with EP in their differential method.  We use lines covering the full range of EPs and, coupled with our more stringent EW upper limit, are able to resolve the issues presented by \citet{Sitnova2015} and remove abundance trends with excitation potential in our line-by-line differential abundance method.

\subsection{Line Lists}
We manually measured the EWs of Fe and $\alpha$-element spectral lines for each  of the nine target metal-poor stars using the IRAF\footnote{IRAF is distributed by the National Optical Astronomy Observatories, which are operated by the  Association of Universities for Research in Astronomy, Inc., under cooperative  agreement with the National Science Foundation.} \emph{splot} routine for Gaussian fits to the line profiles; typical EW uncertainties were found to be $\lesssim$3\%, based on rms scatter around the best-fit profiles.  Several of the lines appeared on adjacent spectral orders in which case the final value used in the analysis was the average of the two measurements.   These repeated lines also provided estimates of our EW measurement uncertainty.

Weak lines were specifically chosen because they are sufficiently sensitive to elemental abundance while also being insensitive to line damping or broadening mechanisms.  For both Fe lines and lines of $\alpha$-elements, a lower limit of $8\,\mathrm{m\AA}$ was established due to the uncertainty in the placement of the continuum level; an upper limit of 85 m\AA\ was enforced to ensure the line behaved well in the linear portion of the curve of growth and wing damping did 
not affect the measurements.  A total of 85 Fe lines were taken from the studies of \citet{McWR1994} and \citet{McW1995} supplemented with lines identified by \citet{NBS61} with line parameters from the Kurucz online database\footnote{http://kurucz.harvard.edu/linelists.html}.  A total of 36 $\alpha$-element lines taken from \citet{Fulb2007} were found to be suitable in the range of EW values.  A final list of the measured EWs for Fe and $\alpha$-elements are provided in Tables~\ref{table:FeLines} and \ref{table:AlphaLines}, respectively.  The line EWs for our standard stars, the Sun and HIP\,66815, are listed separately in Table~\ref{table:ew66815}.

\begin{table*}[t]
\centering
\begin{threeparttable}
\caption{Fe Line List \label{table:FeLines}}
\begin{tabular}{l r r r r r r r r r r r}
\tableline\tableline
Ion & Wavelength & EP & \multicolumn{8}{c}{EW (m\AA )} \\
& (\AA) & (eV) & 46120 & 54639 & 80679 & 87062 & 87788 & 98492 & 103269 & 106924 & 108200\\
\tableline
 Fe~I & 4447.72 & 2.22 &    ...  &    ...  &   55.7  &   53.2  &    ...  &   81.0  &   72.2  &   60.1  &    ...  \\           
 Fe~I & 4469.38 & 3.65 &    ...  &    ...  &   20.3  &   33.0  &    ...  &   56.6  &   42.0  &   27.7  &    ...  \\           
 Fe~I & 4489.74 & 0.12 &    ...  &    ...  &   41.7  &   27.7  &    ...  &   59.7  &   51.9  &   44.7  &    ...  \\           
 Fe~I & 4525.14 & 3.60 &    ...  &    ...  &   15.4  &   29.2  &    ...  &   55.2  &   37.3  &   23.5  &    ...  \\           
 Fe~I & 4736.77 & 3.21 &    ...  &    ...  &   30.2  &   40.4  &    ...  &   63.9  &   54.2  &   39.9  &    ...  \\           
 Fe~I & 4938.81 & 2.88 &    ...  &   30.4  &   28.6  &   36.0  &   12.7  &   62.7  &   49.9  &   39.0  &   61.9  \\           
 Fe~I & 4939.69 & 0.86 &    ...  &   41.0  &   34.7  &   25.7  &   30.1  &   56.0  &   46.7  &   40.3  &   54.3  \\           
 Fe~I & 5072.08 & 4.28 &    ...  &    ...  &    ...  &   10.6  &    ...  &   23.0  &   23.0  &    ...  &    ...  \\           
 Fe~I & 5079.74 & 0.99 &    ...  &   36.4  &   34.1  &   23.8  &   22.3  &   58.3  &   45.3  &   40.0  &   54.7  \\           
 Fe~I & 5083.34 & 0.96 &    ...  &   53.0  &   47.0  &   38.2  &   36.6  &   68.7  &   59.2  &   53.3  &   65.7  \\           
\tableline
\end{tabular}
\begin{tablenotes}
\item{This table is published in its entirety in the electronic edition of the {\it Astrophysical Journal.} A portion is shown here for guidance regarding its form and content.}
\end{tablenotes}
\end{threeparttable}
\end{table*}

\begin{table*}[t]
\centering
\begin{threeparttable}
\caption{Alpha-Element Line List \label{table:AlphaLines}}
\begin{tabular}{l r r r r r r r r r r r}
\tableline\tableline
Ion & Wavelength & EP & \multicolumn{8}{c}{EW (m\AA )} \\
& (\AA) & (eV) & 46120 & 54639 & 80679 & 87062 & 87788 & 98492 & 103269 & 106924 & 108200\\
\tableline
CaI & 4283.01 & 1.89 &    ...  &    ...  &   64.7  &   63.3  &   53.3  &   90.8  &    ...  &   74.1  &    ...  \\           
CaI & 4318.65 & 1.90 &    ...  &    ...  &   51.0  &   64.6  &   30.5  &   81.8  &   69.6  &   62.5  &    ...  \\           
CaI & 4425.44 & 1.88 &    ...  &    ...  &   50.3  &   57.4  &   28.9  &   79.9  &   74.8  &   60.4  &    ...  \\           
CaI & 4435.68 & 1.89 &    ...  &    ...  &   39.2  &   51.1  &   19.7  &   74.5  &   61.6  &   52.3  &    ...  \\           
CaI & 4454.78 & 1.90 &    ...  &    ...  &    ...  &    ...  &   65.3  &  114.7  &    ...  &    ...  &    ...  \\           
CaI & 5265.56 & 2.52 &   33.0  &   31.7  &   24.8  &   40.3  &   14.6  &   63.9  &   46.5  &   38.6  &   56.7  \\           
CaI & 5349.47 & 2.71 &   12.9  &   13.2  &   12.7  &   23.7  &     7.3  &   43.4  &   27.4  &   19.4  &   28.9  \\           
CaI & 5512.98 & 2.93 &    8.7  &    5.4  &    5.6  &   13.2  &    ...  &   29.0  &   14.6  &   12.7  &   14.6  \\           
CaI & 5581.97 & 2.52 &   12.4  &   14.7  &   12.7  &   21.3  &    ...  &   36.5  &   26.5  &   18.3  &   32.0  \\           
CaI & 5588.75 & 2.53 &   51.2  &   58.4  &   52.1  &   52.5  &   28.7  &   78.6  &   72.0  &   63.6  &    85.6  \\            
\tableline
\end{tabular}
\begin{tablenotes}
\item{This table is published in its entirety in the electronic edition of the {\it Astrophysical Journal.} A portion is shown here for guidance regarding its form and content.}
\end{tablenotes}
\end{threeparttable}
\vspace{0.25cm}
\end{table*}

\begin{table}[h!]
\centering
\begin{threeparttable}
\caption{Solar and HIP\,66815 Line Equivalent Widths \label{table:ew66815}}
\begin{tabular}{l r r r r}
\tableline\tableline
Ion & Wavelength & EP & \multicolumn{2}{c}{EW (m\AA )} \\
& (\AA) & (eV) & HIP\,66815 & Sun\\
\tableline
O~I & 7771.944   & 9.146 & 59.3   & 72.6 \\
O~I & 7774.166   & 9.146 & 43.4   & 63.0 \\
O~I & 7775.388   & 9.146 & 37.2   & 49.0 \\
Na~I & 6154.226  &  2.102 &  9.6  &  42.0 \\
Na~I & 6160.747  &  2.104 & 16.8  &  55.4 \\
Mg~I & 4730.029  &  4.346 & 25.1  &  69.6 \\
Mg~I & 5711.088  &  4.346 & 67.4  & 112.5 \\
Mg~I & 6318.717  &  5.108 & 17.6  &  39.4 \\
Mg~I & 6319.237  &  5.108 &  8.9  &  25.5 \\
Mg~I & 6319.495  &  5.108 &  3.1  &  11.5 \\
Mg~I & 8213.013  &  5.753 & 68.0  & 125.2 \\
Mg~I & 8712.690  &  5.932 & 10.2  &  58.9 \\
Mg~I & 8717.820  &  5.933 & 68.0  & 103.0 \\
\tableline
\end{tabular}
\begin{tablenotes}
\item{This table is published in its entirety in the electronic edition of the {\it Astrophysical Journal.} A portion is shown here for guidance regarding its form and content.}
\end{tablenotes}
\end{threeparttable}
\end{table}

\subsection{Stellar Parameters and Atmospheres}\label{sec:StellarParameters}
The abundance indicated from the EW of each absorption line was computed using the \textit{abfind} driver of the 2014 version of the spectrum synthesis program MOOG \citep{Sned1973}.  This analysis required the use of stellar atmospheres generated from the Kurucz grid of LTE models, with updated opacity distributions from Castelli \& Kurucz (2004, henceforth Kurucz models)\footnote{http://kurucz.harvard.edu/grids}.  For the Sun we employed the scaled-solar, ODFNEW, Kurucz model.  Since lines in the spectra of our metal-poor halo stars indicated $\alpha$-element (e.g., O, Mg, Si, Ca, Ti) enhancements, near $+$0.4 dex, typical for their low metallicity, \citep[e.g.,][]{Wall1962, Conti1967}, we employed for those the alpha-enhanced, AODFNEW, Kurucz models.

We note that the Kurucz metallicity parameter, indicated by [M/H], is equal to the adopted [Fe/H] (and not the mass fraction of metals).  For ODFNEW models the solar abundance distribution was scaled by [M/H], while for AODFNEW models the same scaling was applied, but the alpha-elements (O, Mg, Si, S, Ca, Ti) were increased by an additional $+$0.4 dex.

The model atmosphere effective temperature, T$_{\rm eff}$, for each program star was selected so that the differential abundance (ultimately relative to the Sun) derived from our Fe~I lines was independent of line EP.  Since this effective temperature estimate relies on the Saha-Boltzmann excitation of Fe~I we call it the excitation temperature, or T$_{\rm ex}$.  The use of excitation temperatures offers the advantage that they are independent of reddening.  Furthermore, the average differential Fe~I abundance, relative to the Sun, provides our choice of the model atmosphere metallicity. 

Whilst the final stellar model temperatures employed in our abundance analysis  were determined from excitation equilibrium of Fe~I, the initial guess for effective temperatures was based on the (V$-$K) photometric colors for our stars, listed in the SIMBAD database \citep{Deemin1961,Hog2000,Merm1986,Sandage1964}.  Clearly, it is of interest to compare the excitation temperatures, based on the spectra, with the photometric T$_{\rm eff}$ values, based on calibrations of broad-band colors.

Although the majority of the target stars are not significantly affected by reddening, it was noticed that HIP\,87062 and HIP\,98492 showed strong interstellar Na I (5890.0, 5895.9\,\AA) lines requiring a reddening correction to the photometric colors.  Of the remaining program stars, HIP\,108200 showed some weak Na~I interstellar absorption.  

\citet{MunZwit1997} have calibrated a relation between interstellar Na I EWs and reddening that is most sensitive between $0.0\leq E_{B-V}\leq0.4$ with an accuracy of 0.05 mag to 0.15 mag depending on the separation of the stellar and interstellar lines, EW values, and quality of the measurement.  As our data are of such high resolution it should not be the case that the stellar and interstellar lines would be blended and the largest reddening found in our stars is still rather low compared to the range covered by \citet{MunZwit1997} we take 0.05 mag as the reddening uncertainty of our three reddened stars.  The Na I EWs and corresponding reddening factors are provided in Table~\ref{table:Red}.

\begin{table}[h!]
\centering
\caption{Reddening Parameters \label{table:Red}}
\begin{tabular}{l r r r}
\tableline\tableline
HIP\,ID & Na I m\AA\ (D1) & Na I m\AA\ (D2) & $E_{B-V}$\\
\tableline
46120 & 12.0 & 13.0 & 0.00\\
54639 & 11.2 & 7.6: & 0.00\\
80679 & 0.0 & 0.0 & 0.00\\
87062 & 184.0 & 159.0 & 0.06\\
87788 & 0.0 & 0.0 & 0.00\\
98492 & 271.1 & 172.8 & 0.11\\
103269 & 0.0 & 0.0 & 0.00\\
106924 & 0.0 & 0.0 & 0.00\\
108200 & 31.6 & 17.5 & 0.02\\
\tableline
\end{tabular}
\vspace{0.25cm}
\end{table}

These reddening values were applied where necessary to the photometric colors prior to determining the effective temperatures of the stars using color-temperature relations.  We use the \citet{Wink1997} corrections to convert E(B-V) to E(V-K); as a result, the uncertainty of T$_{\mathrm{eff}}$ for the stars without any reddening is $\sim$30 K.  The uncertainty of T$_{\mathrm{eff}}$ for the three stars affected by reddening are listed individually in Table~\ref{table:AtmPar}.  

Because of the metallicity-dependence of the photometric color-temperature relations (redder colors occur at higher metallicity but also with lower temperature), it is necessary to determine both T$_{\rm eff}$ and [Fe/H] iteratively, until both are consistent with the colors and spectra.

\begin{table*}[t]
\centering
\caption{Stellar atmosphere parameters and a comparison of T$_{\rm eff}$
derived from RM05 color--temperature calibrations with spectroscopic values \label{table:AtmPar}}
\begin{tabular}{l c c l |c c c c| c c c c}
\tableline\tableline
 & & & & \multicolumn{3}{c}{Non-Ionization Equilibrium} & & \multicolumn{3}{c}{Ionization Equilibrium}\\
\cline{5-7}\cline{8-11}
\tableline
HIP\,ID & V & (V-K)$_0$ & T$_{eff}$ (K) & T$_{\rm ex}$ (K) & log\,g (dex) & $\xi$ (km/s) & & T$_{\rm ex}$ (K) & log\,g (dex) & $\xi$ (km/s)\\
\tableline
  46120 & 10.10 & 1.64 & 5533 & 5100 & 4.64 & 1.30 && 5250 & 4.20 & 1.60\\
  54639 & 11.41 & 2.06 & 4923 & 4850 & 4.72 & 1.40 && 4900 & 4.30 & 1.70\\
  80679 & 11.29 & 1.89 & 5146 & 4950 & 4.73 & 1.70 && 4950 & 4.40 & 1.30\\
  87062 & 10.59 & 1.65 & $5521\pm68$ & 5800 & 4.42 & 1.70 && 6000 & 4.20 & 1.70\\
  87788 & 11.29 & 1.83 & 5284 & 4900 & 4.71 & 1.70 && 5000 & 4.30 & 1.60\\
  98492 & 11.57 & 1.74 & $5385\pm75$ & 5500 & 4.17 & 1.40 && 5500 & 3.80 & 1.40\\
103269 & 10.33 & 1.72 & 5410 & 5300 & 4.64 & 1.50 && 5325 & 4.20 & 1.50\\
106924 & 10.39 & 1.82 & 5254 & 5125 & 4.66 & 1.70 && 5150 & 4.20 & 1.50\\
108200 & 11.03 & 1.87 & $5200\pm66$ & 5175 & 4.68 & 1.75 && 5175 & 4.60 & 1.50\\
\tableline
\end{tabular}
\end{table*}

The photometric T$_{\rm eff}$ calibration by \citet[henceforth RM05]{RamMel2005} for our six stars with zero reddening are, on average, $\sim$220 K hotter than the spectroscopic T$_{\rm ex}$, as indicated in column 2 of Table~\ref{table:AtmPar}.  Because of this large temperature difference, if we employ the photometric V$-$K T$_{\rm eff}$ for the model atmospheres we find a steep slope of Fe~I abundance with line EP, leading to significant uncertainty of Fe~I abundance.  The photometric color--T$_{\rm eff}$ calibrations of \citet[henceforth Cas10]{Cas2010} result in even hotter stellar temperatures than RM05, some 415\,K higher than our spectroscopic T$_{\rm ex}$, and give steeper slopes of Fe~I abundance with EP.  

\begin{figure*}[t]
\centering
\includegraphics[width=1.0\textwidth]{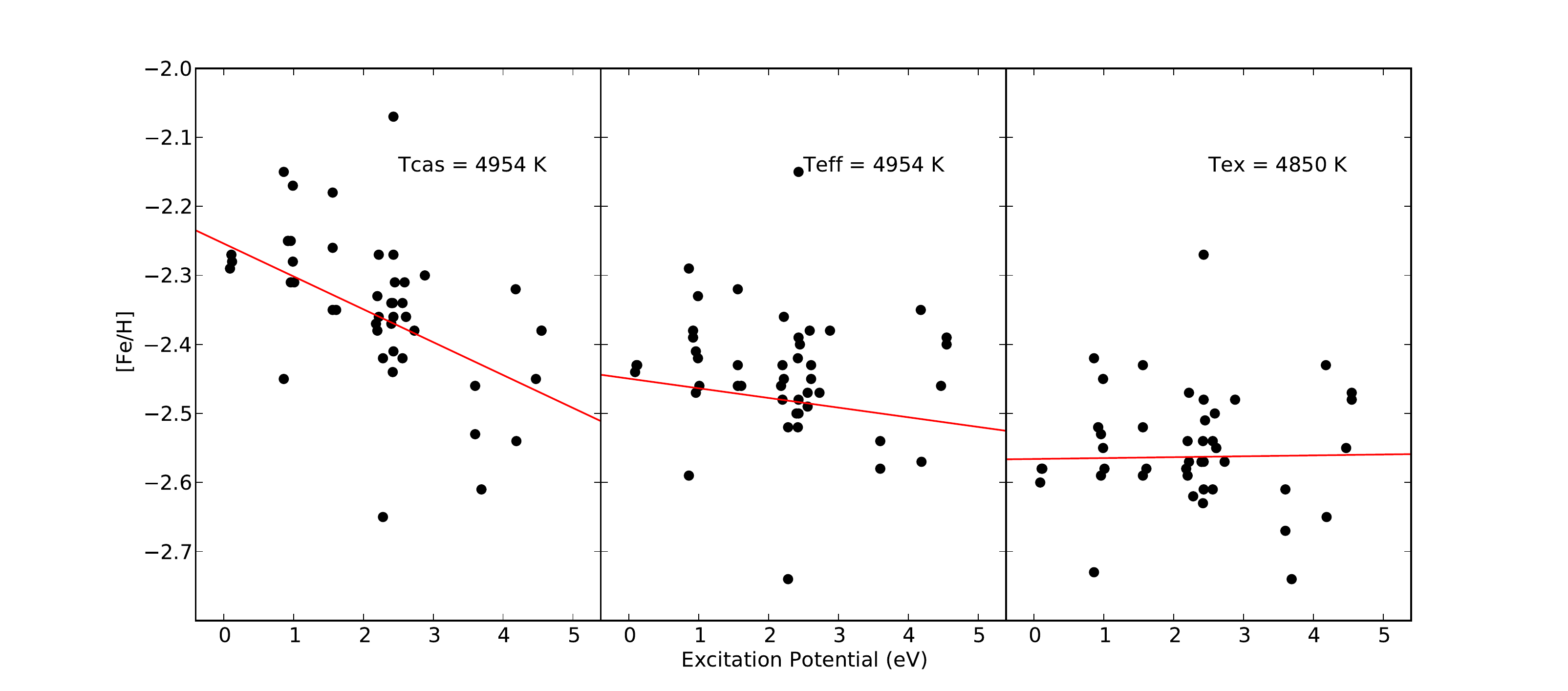}
\caption{Abundance trends with EP for HIP\,54639 based on choice of color-temperature relation.  \emph{Left}: T$_{\mathrm{eff}}$ derived with Cas10 color-temperature relation.  \emph{Center}: T$_{\mathrm{eff}}$ determined with RM05 color-temperature relation.  \emph{Right}: T$_{\mathrm{ex}}$ derived by removing trend in abundance with EP.  The large disagreement between photometric and excitation temperatures is clearly evident from the slopes of the left and center plots.  The abundances derived using these photometric temperatures would depend heavily on line EP, whereas the excitation temperature provides abundances independent of line EP. \label{figure:CompareTemps}}
\vspace{0.25cm}
\end{figure*}

Figure~\ref{figure:CompareTemps} shows the [Fe~I/H] abundance versus EP plot for HIP\,54639, which shows the \emph{smallest} difference between photometric and excitation temperatures, for three different model atmosphere temperatures: the Cas10 color-temperature, the RM05 color-temperature, and the spectroscopic T$_{\rm ex}$.  The plots in Figure~\ref{figure:CompareTemps} demonstrate an enormous disagreement between temperatures based on photometric colors and those derived from the excitation of Fe~I lines.

Given that the 1$\sigma$ uncertainty in the determination of the spectroscopic temperatures is $\sim$25 K, one might ask whether the uncertainty in the reddening alone could account for this difference.  A 0.01\,mag change in $\mathrm{E}(B-V)$ would result in a 12\,K temperature shift.  Therefore, with the average discrepancy between the spectroscopic and photometric temperatures, reddening corrections of $\sim$0.1\,mag would need to be applied to account for the difference.  These values are too high given the small uncertainty in the reddening correction.  To be consistent, we have taken the excitation temperature as the final temperature used throughout the rest of this analysis.

Along the same lines of determining a spectroscopic excitation temperature, we find spectroscopic microturbulent velocity parameters, $\xi$, and surface gravities of our program stars.  For the microturbulent velocity, we followed the usual procedure \citep[e.g.][]{SE1934,Garz1969} of demanding that Fe~I line abundances are independent of EW.  Our reliance on weak, unsaturated, lines for the Fe~I abundances of our program stars provides relative insensitivity to uncertainties on the exact value of $\xi$.  For the surface gravities, we make an initial estimate based on 12.0 Gyr DSEP isochrones with a helium fraction Y$ = 0.245+1.5Z$, $[\alpha/\mathrm{Fe}] = +0.4$, and initial [Fe/H] = $-$2.0 and adjust this initial estimate until ionization equilibrium between [Fe\,I/H] and [Fe\,II/H] is achieved. 

At the low metallicity of our program stars the [$\alpha$/Fe] ratios are expected to be enhanced by approximately $+$0.4 dex \citep[e.g.][]{Wall1962,Conti1967}.  For this reason, we have employed the alpha-enhanced, AODFNEW, Kurucz model atmospheres in our abundance analysis.  However, in order to check that our stars actually possess halo-like $\alpha$-enhancements, we have measured the EWs of a number of lines from $\alpha$-elements, and from those derive [$\alpha$/Fe] abundance ratios in a line-by-line differential analysis.

\subsection{Differential Analysis of Standard Star HIP\,66815}\label{sec:hip66815}
In order to use standard star HIP\,66815 in our line-by-line differential abundance analysis, we first need to determine its atmosphere parameters, [Fe~I/H], [Fe~II/H] and [$\alpha$/Fe] values.

\subsubsection{Solar Atmosphere Parameters}
For our 1D LTE differential abundance analysis of HIP\,66815 we require physical parameters for the solar atmosphere, including the microturbulent velocity.  Here we adopt the solar T$_{\rm eff}$ of 5777\,K and log\,g of 4.4377\,dex given by \citet{AQ2000}.  We have obtained the microturbulent velocity for the Sun in an absolute abundance analysis, using the Fe~I line EWs and log\,gf values of \citet{Black1995a,Black1995b} and \citet{Holweger1995}, as listed in \cite{Asplund2000}.  While differences in the log\,gf scales of the Holweger and Blackwell groups gave slightly different solar Fe abundances, within each log\,gf scale the Fe abundance trend with EP was flat, indicating agreement with the solar effective temperature of 5777\,K.  A least squares fit of Fe~I EWs versus absolute solar abundances, derived using the Asplund lines and Holweger log\,gf values, gave zero slope for a microturbulent velocity parameter of 0.99\,km/s, with a formal 1$\sigma$ uncertainty of $\pm$0.07\,km/s.  Unfortunately, systematic errors made it impossible to simply include solar Fe~I abundances found from the lines with Blackwell log\,gf values.  Therefore, we have adopted a solar microturbulent velocity parameter rounded to 1.0\,km/s.  

\subsubsection{HIP\,66815 Atmosphere Parameters}
Our initial physical atmosphere parameters for HIP\,66815 were based on the \emph{Hipparcos} parallax of 18.43$\pm$1.07 mas \citep{Leeu2007}, V=8.83, and 2MASS V$-$K and V$-$J colors of 1.44 and 1.10, respectively, suggesting T$_{\rm eff}$=5864 K and log\,g=4.50 dex using the RM05 color-temperature relations, and assuming a mass near 0.76 M$_{\odot}$, inferred from a 12~Gyr DSEP isochrone, appropriate for an old metal-poor halo star.

\begin{figure*}[t]
\centering
\includegraphics[width=0.75\textwidth]{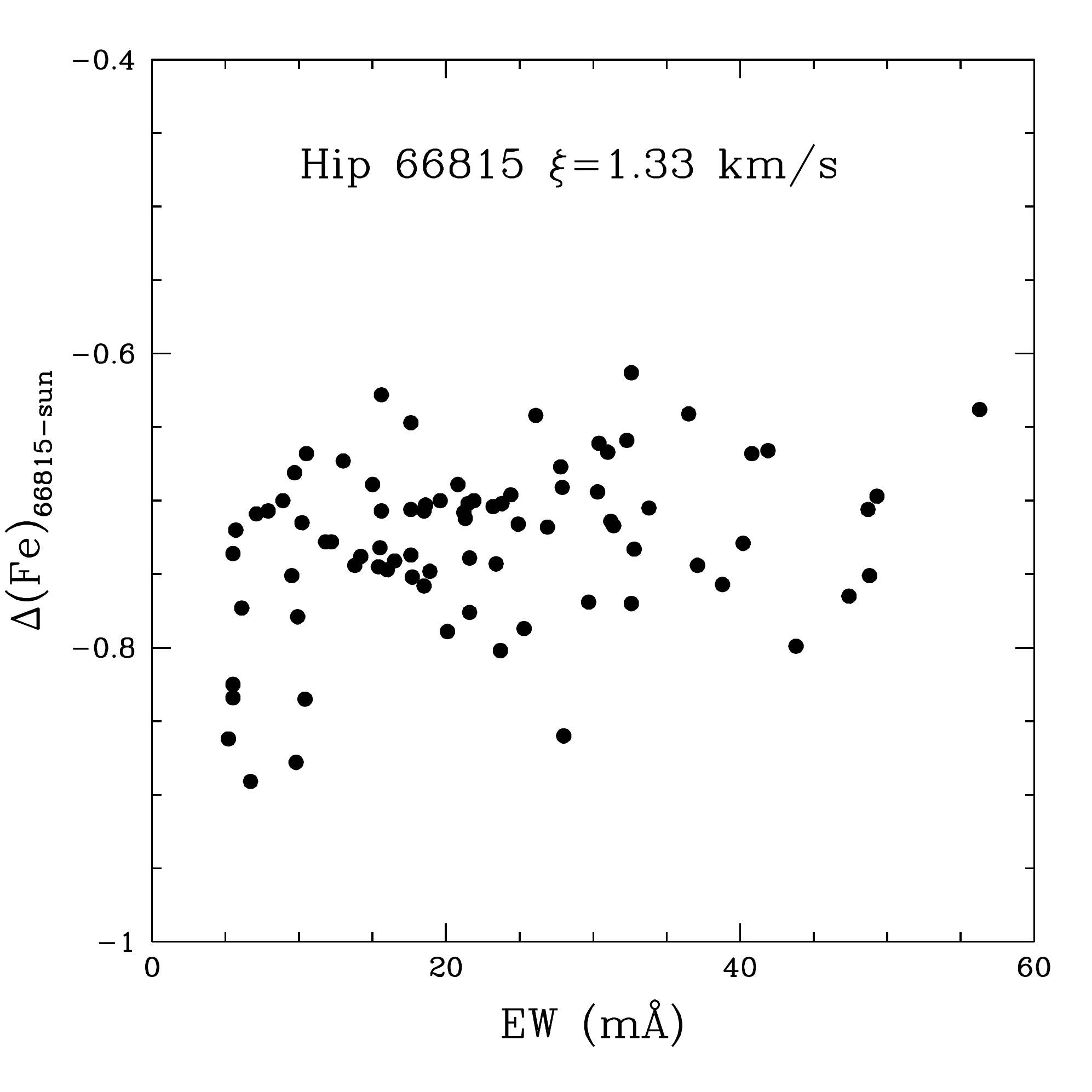}
\caption{A plot of $\Delta\varepsilon$(Fe~I) versus EW in HIP\,66815 for a microturbulent velocity parameter of $\xi$=1.33 km/s, showing iron abundance independent of EW. The flat trend of abundance with EW indicates that the appropriate model atmosphere microturbulent velocity is obtained. \label{fig:ewab66815}}
\vspace{0.25cm}
\end{figure*}

\begin{figure*}[t]
\centering
\includegraphics[width=0.75\textwidth]{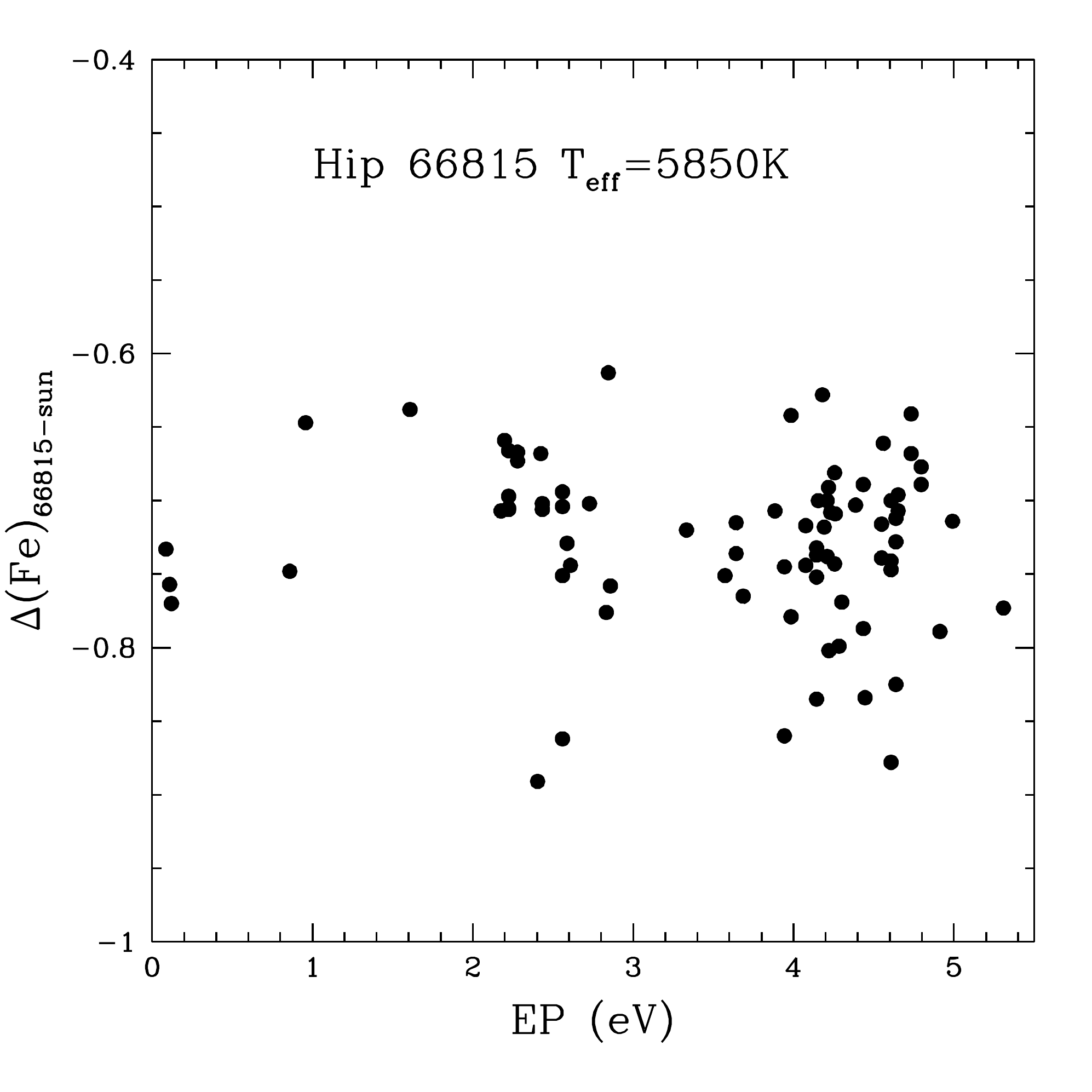}
\caption{A plot of $\Delta\varepsilon$(Fe~I) versus EP in HIP\,66815 for T$_{\rm eff}$=5850 K.  The flat trend of abundance with EP indicates that the appropriate model atmosphere temperature is obtained. \label{fig:epab66815}}
\vspace{0.25cm}
\end{figure*}

Notably, the color-temperature calibration of Cas10 would have indicated a higher temperature of 5974 K, for HIP\,66815.  We have also determined an excitation temperature for HIP\,66815 by finding the model atmosphere yielding Fe~I abundances independent of line EP.  We only consider lines in HIP\,66815 with EW less than 85\,m\AA\ in order to avoid systematic errors arising from difficulties in measuring damped wings of the same lines in the Sun, typically near 120--130\,m\AA.  Accordingly, with $\alpha$-enhanced Kurucz model atmospheres and 81 Fe~I lines, we found an excitation temperature of 5822\,K, with a 1$\sigma$ uncertainty of 31\,K, and a microturbulent velocity parameter of 1.33 km/s (see Figures~\ref{fig:ewab66815} and \ref{fig:epab66815}).

These parameters are relatively close to those found by \citet{KM2008, KM2011}, who obtained T$_{\rm eff}$=5812 K, log\,g=4.41 dex, [Fe/H]=$-$0.76 dex, and $\xi$=1.13 km/s.  This star was also analyzed by \citet{Fulb2000} and \citet{Sobeck2006}.  \citet{Fulb2000} derived an excitation temperature, based on laboratory gf values and found T$_{\rm eff}$=5875K, with a 1$\sigma$ uncertainty of 40K; a spectroscopic log\,g of 4.50 dex, was chosen by \citet{Fulb2000} in order to force ionization equilibrium of iron, [Fe~I/H]=[Fe II/H], again based on laboratory gf values.  \citet{Sobeck2006} claimed to employ the \citet{Fulb2000} atmosphere parameters and computed [Fe/H]=$-$0.63 dex, but reported [Fe/H]=$-$0.64 dex for the \citet{Fulb2000} value; this conflicts with the on-line data of \citet{Fulb2000}, which indicates $\mathrm{[Fe/H]}=-$0.5\,dex.  We suspect a typographical error in the on-line table entry for [Fe/H] in \citet{Fulb2000}.  \citet{Fulb2000} found a microturbulent velocity of 0.95\,km/s for HIP\,66815, lower than that found here. 

We are encouraged by the good agreement between the excitation temperature and photometric temperatures, indicated by V$-$K and V$-$J photometry, based on the RM05 temperature-color relations. Therefore, we adopt the average of these three temperatures (from V$-$K, V$-$J, and T$_{\rm ex}$), at T$_{\rm eff}$=5850 K and $1\sigma$=24 K.  We reject the photometric temperature indicated by the Cas10 calibration, as this is 152 K higher than our excitation temperature, nearly 5$\sigma$ beyond our estimated uncertainties.  Such a high temperature would impose a strong slope of Fe~I abundance with line EP and make it difficult to determine [Fe/H].  Interestingly, \citet{Huang2015} found that the Cas10 T$_{\rm eff}$ values were hotter than direct measurements by 131~K in a comparison of 69 dwarf stars.  On the other hand, a disagreement is implied with \citet{Cas2011}, who claim no significant shift between their color-temperature scale and excitation temperatures.

The final adopted T$_{\rm eff}$ of 5850K and $\mathrm{[Fe/H]}=-0.72$\,dex, consistent with the spectra, when combined with the 2MASS K-band magnitude and \emph{Hipparcos} parallax are best fit with an 11 Gyr DSEP isochrone and $\mathrm{log }g=4.48$\,dex.

For an alpha-enhanced Kurucz model for HIP\,66815, with T$_{\rm eff}$=5850 K, log\,g=4.50 dex, and $\mathrm{[Fe/H]}=-0.75$\,dex, the 81 Fe~I line EWs give [Fe/H]=$-$0.723 dex, with 1$\sigma$ dispersion of 0.055 dex, providing a formal random error on the mean of 0.006 dex.  If a temperature of 5822 K is employed the [Fe/H] is lowered by an additional 0.02 dex.  Based on 11 Fe~II lines this model for HIP\,66815 gives an average [Fe/H]=$-$0.587 dex, with 1$\sigma$ dispersion of 0.032 dex, indicating an error on the mean of 0.010 dex.  Thus, our analysis of HIP\,66815 has not achieved ionization balance.  

Various factors may have contributed to the 0.14 dex difference between [Fe~I/H] and [Fe~II/H]: an over-estimated model atmosphere electron number density (N$_e$); an inexplicably low log\,g, by $\sim$0.38 dex, which could be due to a mass 60\% lower than indicated by the isochrones, or to an under-estimated luminosity, due to a $\sim$55\% error in the measured parallax, or an actual stellar radius 55\% larger than the DSEP isochrones; a strong non-LTE over-ionization of iron, leading to low [Fe~I/H]; or an under-estimated T$_{\rm eff}$ by $\sim$150 K, perhaps due to non-LTE or 3D effects, also leading to low [Fe~I/H].

Of these possibilities, an overestimated N$_e$ could result from an inappropriate [$\alpha$/Fe] model atmosphere parameter.  In Table~\ref{table:ab66815}, we show the results of our differential chemical abundance analysis of HIP\,66815, relative to the Sun, based on the EWs listed in Table~\ref{table:ew66815}.  Our measurements indicate  [$\alpha$/Fe] ratios for HIP\,66815 that are relatively low for its metallicity, near $+$0.10 dex for the electron donors Mg, Si, Ca and Ti, but $+$0.28 dex for O.  This is much less than the $+$0.4 dex enhancements employed in the Kurucz model atmospheres.  Thus, a solar composition model atmosphere is more appropriate for HIP\,66815 than an alpha-enhanced model; use of the solar composition model atmosphere for HIP\,66815 gives [Fe~I/H]=$-$0.718 (rms dispersion 0.055 dex) and [Fe~II/H]=$-$0.627 (rms dispersion 0.030 dex).  Interpolating between models for [$\alpha$/Fe]=$+$0.10 dex, which is appropriate for the electron donor alpha-elements, we obtain [Fe~II/H]=$-$0.617 dex, as the best estimate for HIP\,66815; thus, the Fe ionization imbalance is reduced to 0.10 dex. 

\begin{table}[h!]
\centering
\caption{HIP\,66815 Chemical Composition\tablenotemark{1} \label{table:ab66815}}
\begin{tabular}{l r r r}
\tableline\tableline
Ion & [X/Fe]\tablenotemark{2} &$\sigma$\tablenotemark{3} &N$_{\rm lines}$\\
\tableline
Fe~I  & $-$0.72 &   0.06   &    81 \\
Fe~II & $-$0.59 &   0.03   &    11 \\
O~I\tablenotemark{d}   & $+$0.28 &   0.07   &     3 \\
Na~I  & $-$0.05 &   0.04   &     2 \\
Mg~I  & $+$0.12 &   0.18   &     8 \\
Al~I  & $-$0.03 &   0.04   &     5 \\
Si~I  & $+$0.10 &   0.03   &     9 \\
Ca~I  & $+$0.09 &   0.05   &     7 \\
Ti~I  & $+$0.09 &   0.12   &     4 \\
Ti~II\tablenotemark{4} & $+$0.09 &   ...    &     1 \\
\tableline
\tablenotetext{1}{Differential LTE abundances, using an AODFNEW Kurucz model atmosphere}
\tablenotetext{2}{For Fe~I and Fe~II [Fe/H] is indicated}
\tablenotetext{3}{Random error on the mean [X/Fe] values are given by $\sigma$/$\sqrt{(N-1)}$}
\tablenotetext{4}{[Fe~II/H] used to compute [X/Fe] for O~I and Ti~II}
\end{tabular}
\end{table}

Use of the MARCS model atmospheres \citep{Gust2008} gives the same abundances as the Kurucz model, with differences less than 0.01 dex.  While HIP\,66815 appears to be relatively $\alpha$-deficient for its [Fe/H], [O/Mg] versus [O/H] and [Mg/H] are completely consistent with normal trends throughout the Milky Way components \citep[e.g.,][]{McW2008,Bensby2005,Fulb2007}. 

Regarding over-ionization of Fe~I by non-LTE effects in HIP\,66815, the calculations of \citet{Lind2012} show this is very small, near $\sim$0.03 dex, but negligible in a differential abundance of HIP\,66815 relative to the Sun.  Furthermore, the \citet{Lind2012} calculations also showed a negligible non-LTE effect (less than 10 K) on the derived excitation temperature.  Negligible non-LTE differential corrections for Fe abundances in HIP\,66815 are supported by the review of \citet{Mash2014} and the discussion in Section~\ref{sec:NLTE}.

Interestingly, the non-LTE calculations of \citet{Berg2012} showed no significant errors in spectroscopically derived 1D-LTE Teff and logg for dwarf stars similar to HIP66815.  The 1D-LTE Teff and logg should give an answer very close to the truth ($\Delta T = 5$ K, $\Delta \mathrm{log}\,g = 0.02$\,dex).  In the same paper, \citet{Berg2012} found an analysis of the 1D spatially averaged 3D (henceforth $\langle$3D$\rangle$) solar model of \citet{Coll2011} indicated non-LTE corrections for Fe~I lines always less than $\sim$0.02 dex (i.e. negligible).  Similar $\langle$3D$\rangle$ models are discussed further in Section~\ref{sec:3Dmodels}.

Notwithstanding uncertainties regarding 3D model atmospheres, none of the effects listed above appears capable of explaining the 0.10 dex Fe ionization imbalance. However, it is possible that a combination of systematic effects might be the cause of the imbalance.  For example, a 30 K, 1$\sigma$, increase in T$_{\rm eff}$ reduces $\Delta$Fe by 0.03\,dex; a 2$\sigma$ decrease in the parallax of HIP\,66815 accounts for a further reduction of 0.04 dex; and a 0.02--0.03 dex non-LTE differential Fe over-ionization effect is not unreasonable.  To aid in understanding the roles of systematic and random errors on the Fe ionization balance, it would be useful to consider Fe~I and Fe~II lines for all our program stars.  Concerns about increased 3D effects at low metallicity would also be addressed with [Fe/H] measured from Fe~II lines.

Until the ionization imbalance is resolved we have two metallicity scales for HIP\,66815: 81 Fe~I lines favor [Fe/H]=$-$0.72 dex (rms 0.06 dex), while 11 Fe~II lines favor [Fe/H]=$-$0.62 dex (rms 0.03 dex).

\section{Differential LTE Abundance Results for the Program Stars}\label{sec:DiffAbund}
Similar to the analysis of HIP\,66815 relative the Sun we performed a line-by-line differential abundance analysis of our program stars relative to HIP\,66815, using the EWs listed in Tables~\ref{table:FeLines} and \ref{table:AlphaLines} and the atmosphere parameters listed in Table~\ref{table:AtmPar}. The abundance ratios were normalized to the solar scale using our HIP\,66815 chemical composition, given in Table~\ref{table:ab66815}, with the final LTE [Fe/H] results listed in Table~\ref{table:FeResults}.  We remind the reader that $\alpha$-enhanced, AODFNEW, Kurucz models were employed for both the program stars and the standard, HIP\,66815, in the differential analysis.

Table~\ref{table:FeResults} lists our LTE [Fe~I/H] and [Fe~II/H] results.  As will be discussed in Section~\ref{sec:IonEq}, the use of isochrone gravities resulted in inconsistent Fe~I and Fe~II abundances for all of our stars.  In order to approach ionization equilibrium, where Fe~I and Fe~II abundances are equal, it was necessary to use a log\,g that was, on average, 0.4\,dex lower than that predicted by the DSEP isochrones.  Because of this, in the following discussion we provide atmospheric parameters and abundances both with and without ionization equilibrium in Table~\ref{table:AtmPar}.  In Tables~\ref{table:FeResults} and \ref{table:aResults}, the subscript IE indicates results for which the model atmosphere log\,g was found by forcing ionization equilibrium.

Table~\ref{table:aResults} lists our LTE [X/Fe] ratios for alpha-elements, O, Mg and Ca.  Due to the low metallicity, we were unable to detect O~I lines in most of the program stars, and many of the Mg~I lines.  Typically, three Ca~I lines were seen in our program stars; however, for our most metal-poor star, HIP\,87788, only one Ca~I line was detected.  For our most $\alpha$-enhanced star, HIP\,98492, we measured five Ca~I lines.  Thus, our measured [$\alpha$/Fe] ratios are strongly weighted by Ca~I lines and our measured [Ca/Fe] ratios (with a mean [Ca/Fe] = 0.34 dex) are consistent with the enhanced [$\alpha$/Fe] ratios of the MW halo \citep[and references therein]{Sneden2004}.

\begin{table*}{t}
\centering
\caption{LTE Fe Abundance Results \label{table:FeResults}} 
\begin{tabular}{r c c c c c c c c c}
\tableline\tableline
& \multicolumn{4}{c}{[Fe~I/H]} && \multicolumn{4}{c}{[Fe~II/H]}\\
\cline{2-5}\cline{7-10}
HIP\,ID & Fe~I & Fe~I$_{\mathrm{IE}}$ & $\sigma$ & N & & Fe~II & Fe~II$_{\mathrm{IE}}$ & $\sigma$ & N\\
\tableline
46120 & -2.33 & -2.22 & 0.13 & 40 & & -1.87 & -2.18 & 0.28 & 3\\
54639 & -2.48 & -2.50 & 0.15 & 48 & & -2.14 & -2.46 & 0.28 & 5\\
80679 & -2.49 & -2.50 & 0.08 & 52 & & -2.30 & -2.46 & 0.21 & 12\\
87062 & -1.70 & -1.56 & 0.15 & 52 & & -1.37 & -1.54 & 0.18 & 15 \\
87788 & -2.69 & -2.66 & 0.10 & 32 & & -2.42 & -2.60 & 0.24 & 7 \\
98492 & -1.40 & -1.40 & 0.08 & 69 && -1.17 & -1.35 & 0.13 & 6 \\
103269 & -1.85 & -1.83 & 0.07 & 66 & & -1.64 & -1.84 & 0.20 & 16 \\
106924 & -2.22 & -2.23 & 0.09 & 60 & & -1.86 & -2.25 & 0.21 & 11\\
108200 & -1.82 & -1.83 & 0.12 & 47 & & -1.75 & -1.79 & 0.12 & 5 \\
\tableline
\end{tabular} 
\end{table*} 

\begin{table*}[t]
\centering
\caption{LTE $\alpha$-element Abundance Results \label{table:aResults}} 
\begin{tabular}{r c c c c c c c c c c c c c c}
\tableline\tableline
& \multicolumn{4}{c}{[O/Fe]} && \multicolumn{4}{c}{[Mg/Fe]} && \multicolumn{4}{c}{[Ca/Fe]}\\
\cline{2-5}\cline{7-10}\cline{12-15}
HIP\,ID & O & O$_{\mathrm{IE}}$ & $\sigma$ & N & & Mg & Mg$_{\mathrm{IE}}$ & $\sigma$ & N & & Ca & Ca$_{\mathrm{IE}}$ & $\sigma$ & N\\
\tableline
46120   &  ... &  ...  & ...  & ... &&  0.12 & 0.08 &  N/A & 1   &&  0.41 & 0.41 & 0.15 & 3 \\
54639   &  ... &  ...  & ...  & ... &&    ... &   ... &  ... & ... &&  0.27 & 0.33 & 0.12 & 3 \\
80679   &  ... &  ...  & ...  & ... &&    ... &   ... &  ... & ... &&  0.35 & 0.38 & 0.13 & 3 \\
87062   & 0.01 & 0.11 & 0.02 &  3  &&  0.54 & 0.47 & 0.50 & 3   &&  0.26 & 0.21 & 0.03 & 3 \\
87788   &  ... &  ...  & ...  & ... &&    ... &   ... &  ... & ... &&  0.78 & 0.79 &  N/A & 1 \\
98492   & 0.69 &  0.71 & 0.08 &  3  &&   0.39 &  0.41 &  N/A & 1   &&  0.19 & 0.20 & 0.05 & 5 \\
103269  & 0.68 &  0.69 & N/A &  1  &&  0.15 & 0.14 & N/A & 1   &&  0.23 & 0.22 & 0.08 & 5 \\
106924  &  ... &  ...  & ...  & ... &&   0.28 &  0.27 & N/A & 1   &&  0.40 & 0.40 & 0.05 & 3 \\
108200  &  ... &  ...  & ...  & ... &&  0.06 & 0.05 &  N/A & 1   && 0.12 & 0.14 & 0.06 & 3 \\
\tableline
\end{tabular}
\vspace{0.25cm}
\end{table*} 

\subsection{Ionization (non-)equilibrium \label{sec:IonEq}}
As can be seen from Table~\ref{table:FeResults}, our isochrone log\,g atmospheres gave [Fe~I/H] abundance results significantly lower than [Fe~II/H] for all of our stars; on average, the Fe~I abundances were 0.27 dex lower than those from Fe~II.  This exceeds the ionization imbalance for HIP\,66815 of 0.10 dex.

If the model atmospheres and spectral synthesis programs replicate stellar atmospheres perfectly then the Fe~I and Fe~II abundances should be the same.  A solution to the ionization imbalance might be sought in non-LTE corrections, since non-LTE effects often over-ionize the minority neutral species, like Fe~I, but leave the dominant, singly ionized species, Fe~II, relatively unaffected, and result in a departure from ionization equilibrium \citep[e.g.][]{AthayLites72}.  However, the non-LTE corrections for Fe~I lines in our program stars are negligible, of order 0.01 dex, based on calculations from the INSPECT\footnote{http://inspect.coolstars19.com/} program, v1.0 (see \citet{Lind2012} and Section~\ref{sec:NLTE}). 

In LTE model atmospheres, the electron number density, $N_e$, controls the strength of lines from the dominant ionization state (e.g., Fe~II in our stars), and thus determines whether ionization equilibrium is obtained.  While N$_e$ is strongly sensitive to the surface gravity, it is also sensitive to other effects, like over-all metallicity, the abundance of low-ionization species, such as $\alpha$-elements, and temperature.  Typically, Fe~II lines are stronger in lower surface gravity models whilst Fe~I line strengths do not change much with log\,g; because of this, it has become common practice to choose a spectroscopic gravity by finding the model atmosphere that results in ionization equilibrium.
      
In this analysis we have modified the model atmosphere gravity in order to approach ionization equilibrium.  This has resulted in an average decrease in log\,g of 0.4 dex from the isochrone values.  In LTE ionization equilibrium, our Fe~II abundances are still, on average, 0.03 dex higher than the Fe~I abundances, quite similar to the scale of the non-LTE abundance corrections.  The non-LTE abundance corrections are discussed in greater detail, for various elements, in the next section.

\subsection{NLTE corrections} \label{sec:NLTE}
Here, we consider the potential corrections, due to non-LTE effects, to be applied to our differential LTE abundances.  Because our very metal-poor program stars lack strong line blanketing at blue and UV wavelengths, the path length of such photons is large.  Thus, UV photons from hot regions of the stellar atmosphere can travel to, and be absorbed in, much cooler regions, and thereby affect the excitation and ionization populations of atomic species.  To compute synthetic spectra in non-LTE requires a detailed consideration of radiative and collisional excitation and de-excitation rates for each level of the atom or ion, and consideration of radiation from the whole atmosphere.  This is computationally much more taxing than the simple LTE  situation where the total population of a species is derived from one atomic transition using the Saha-Boltzmann equation.  In addition, it is unfortunate for the non-LTE case that collisional rates are normally not well known; in particular, the rate of collision with hydrogen atoms is parameterized with a scaling factor applied to the formula of \citet{Drawin1968,Drawin1969}.

Since our differential, line-by-line, abundance analysis was performed relative to the solar spectrum, only the differential non-LTE corrections between the Sun and program stars need to be applied to our LTE abundances.

We first consider non-LTE effects on Fe because we employ Fe~I and Fe~II line abundances to constrain our model atmosphere parameters. We are encouraged by the extensive 1D non-LTE investigation of \citet{Lind2012} showing non-LTE corrections to the Fe~I and Fe~II abundances for our program stars are negligible, with no significant effect on the spectroscopically derived atmospheric parameters.  Specific non-LTE calculations by Bergemann (2011 unpublished; but see also \citealt{BN2014}), showed $\sim$$+$0.06 dex non-LTE abundance corrections for Fe~I lines used here; however, differential to the Sun, the non-LTE Fe~I corrections are +0.02 to +0.03 dex higher than our LTE results.  On the other hand, the non-LTE web tool INSPECT\footnote{http://inspect.coolstars19.com/} program, v1.0 (see \citet{Lind2012}), indicates non-LTE corrections for Fe~I lines used in our program stars of only +0.01 dex.  This magnitude for the non-LTE effect on Fe~I abundances in our stars is supported by the review of non-LTE by \citet{Mash2014}.

Non-LTE corrections for the O~I infrared triplet lines used in this analysis (at 7772, 7774, and 7775\AA) have been computed for an extensive grid of FGK dwarf model atmospheres by \citet{Amarsi2016}.  The results are similar to previous calculations \citep[e.g.,][]{Fabb2009,Sitnova2013}.  These non-LTE corrections increase strongly with temperature and with decreasing log~g.  Typically, the lower levels of the O~I triplet lines are over-populated relative to LTE in our stars, so the LTE abundances are too high.  However, since the non-LTE correction is larger for the Sun than for our, relatively cooler program stars, net positive corrections to the LTE [O/H] results are required for our program stars.  For the three program stars where we were able to measure O~I triplet line EWs, we have interpolated the grid of non-LTE corrections of \citet{Amarsi2016} to the appropriate atmosphere parameters; we also interpolated the grid of non-LTE O~I corrections for the Sun.  Consequently, the non-LTE oxygen abundance corrections for HIP\,87062, HIP\,98492, and HIP\,103269 are $-$0.07, $-$0.13, and $-$0.05\,dex, respectively.  For differential abundance relative to the Sun the solar non-LTE correction, at $-$0.18 dex, must be subtracted.  Thus, the [O/H] ratios of our three program stars should be increased by $+$0.11, $+$0.05, and $+$0.13 dex, respectively.  The non-LTE corrected [O/Fe] ratios must also include the differential non-LTE effect on the Fe abundance, of approximately $+$0.02 dex.  Thus, the LTE [O/Fe] ratios for HIP\,87062, HIP\,98492, and HIP\,103269 should be increased by $+$0.09, $+$0.03 and $+$0.11 dex respectively.

For our Mg~I lines we consult \citet{ZG2000}, who computed non-LTE corrections for a range of stellar parameters, encompassing the parameters similar to those of the stars studied here.  These corrections are sensitive to temperature, with larger corrections in hotter stars.  For the Mg~I lines employed here, the \citet{ZG2000} absolute non-LTE corrections are all near, or below, 0.10 dex; and, the non-LTE correction for the stars near 5200 K is 0.07 dex.  The coolest star in \citet{Gehren2006}, with a temperature of 5070 K has a non-LTE Mg~I correction of only 0.03 dex.

Relative to the Sun, the Mg~I non-LTE effect in our metal-poor dwarf stars is reduced due to the solar non-LTE correction.  Thus, our hotter program stars have a mean non-LTE Mg~I differential correction of 0.07 dex, while for stars at 5200 K the differential correction is only 0.02 dex.

For non-LTE corrections to our Ca~I LTE abundances we used the extensive compilation of \citet{Mash2007}, whose computations cover most of the parameter space of our program stars and includes 16 Ca~I lines, including good overlap with the lines employed here.  While the solar non-LTE correction is small, near 0.02\,dex, corrections for our program stars are all near 0.14\,dex.  Thus, a differential non-LTE correction of +0.12\,dex should be added to our Ca~I abundance values.

\subsection{Abundance Uncertainties}\label{sec:Errors}
In order to estimate systematic abundance errors, due to uncertainties in atmosphere parameters, we performed an error analysis in which we varied the atmospheric parameters by conservative quantities ($T\pm50$\,K; $\mathrm{log }g\pm 0.4$\,dex; $\xi\pm$ 0.1 km/s; [M/H] $\pm$ 0.1 dex) and recomputed abundances in order to determine the effect each parameter may have on the final abundance ratios.  We show the effects on three stars in Table~\ref{table:Errors} listed as the difference of the original abundance from the abundance with the modified parameters.  

\begin{table*}[t]
\centering
\caption{Error Analysis \label{table:Errors}}
\begin{tabular}{c c r r r r r r r r r r r}
\tableline\tableline
HIP\,ID &  & \multicolumn{2}{c}{$\Delta T_{ex}$} && \multicolumn{2}{c}{$\Delta$ log\,g} && \multicolumn{2}{c}{$\Delta\xi$}&& \multicolumn{2}{c}{$\Delta [M/H]$}\\
 \cline{3-4}\cline{6-7}\cline{9-10}\cline{12-13}\\
 & & -50 K & +50 K && -0.4 dex & +0.4 dex && -0.1 km/s & +0.1 km/s && -0.1 dex & +0.1 dex \\
\tableline
 & [Fe~I/H] & -0.06 & 0.04 & & -0.05 & 0.04 & & 0.01 & -0.02 & & 0.01 & -0.01\\
HIP\,87788 & [Fe~II/H] & 0.02 & -0.01 & & -0.19 & 0.18 & & 0.01 & -0.01 & & -0.01 & 0.02\\
 & [Ca/Fe] & 0.03 & -0.04 & & -0.01 & 0.03 & & 0.01 & $<0.01$ & & 0.01 & $<0.01$\\
\tableline
 & [Fe~I/H] & -0.05 & 0.05 & & -0.02 & 0.02 & & 0.01 & -0.01 & & $<0.01$ & $<0.01$\\
 & [Fe~II/H] & 0.02 & 0.02 & & -0.16 & 0.17 & & 0.01 & -0.01 & & -0.01 & 0.02\\
HIP\,103269 & [O/Fe] & 0.04 & -0.04 & & -0.02 & $<0.01$ & & -0.01 & 0.01 & & 0.01 & -0.02\\
 & [Mg/Fe] & 0.03 & -0.02 & & 0.01 & $<0.01$ & & $<0.01$ & 0.01 & & 0.01 & $<0.01$\\
 & [Ca/Fe] & 0.02 & -0.02 & & 0.01 & -0.01 & & -0.01 & 0.01 & & 0.01 & $<0.01$\\
\tableline
 & [Fe~I/H] & -0.05 & 0.04 & & -0.04 & 0.04 & & $<0.01$ & -0.01 & & $<0.01$ & -0.01\\
 & [Fe~II/H] & 0.01 & -0.01 & & -0.19 & 0.18 & & 0.01 & -0.01 & & -0.01 & 0.01\\
HIP\,106924 & [Mg/Fe] & 0.05 & -0.01 & & 0.02 & $<0.01$ & & $<0.01$ & 0.01 & & 0.01 & 0.01\\
& [Ca/Fe] & 0.01 & -0.01 & & 0.02 & -0.02 & & $<0.01$ & 0.01 & & $<0.01$ & $<0.01$\\
\tableline
\end{tabular}
\end{table*}

As indicated in Section~\ref{sec:StellarParameters} for star HIP\,54639, a 1$\sigma$ 50\,K effective temperature uncertainty in the error analysis is conservative for the stars in our sample.  By using the spectroscopic temperatures and requiring ionization equilibrium to determine surface gravities we find the spectroscopic surface gravity to be 0.4 dex lower than that predicted from the isochrones and assume a conservative offset of 0.4 dex as our maximum log\,g uncertainty.  We also assume a possible microturbulent velocity uncertainty of 0.05 km/s.

The results provided in Table~\ref{table:Errors} may be used to estimate the systematic abundance uncertainties.  We add in quadrature the average uncertainty from each parameter in Table~\ref{table:Errors} to determine the average systematic uncertainty.  The actual uncertainties are likely to be smaller due to the covariance among parameters \citep{McW1995}.  We find $\sigma_{\mathrm{Fe I}}=0.07$ dex and $\sigma_{\mathrm{Fe II}} = 0.18$ dex for [Fe I/H] and [Fe II/H], respectively. The size of the $1\sigma$ uncertainty on [Fe II/H] is due mainly to the uncertainty incurred by the reduction in log\,g by $\sim$0.4 dex to obtain ionization equilibrium.  The $\alpha$-element abundance uncertainties are $\sigma_\mathrm{OI} = 0.03$ dex, $\sigma_{\mathrm{Mg}} = 0.04$ dex, and $\sigma_{\mathrm{Ca}} = 0.05$ dex, where the high uncertainty for [Mg/Fe] stems from the lack of available line measurements.

The stellar abundances for each star, including the total $1\sigma$ uncertainties, are provided in Table~\ref{table:IEresults}.  The uncertainties are found by adding quadrature the average systematic uncertainty and the rms uncertainty of the abundance for a given star.  For stellar abundances that are measured from a single line, the rms abundance uncertainty is the average of rms abundance uncertainties for that species from stars with multiple lines measured (including HIP\,66815).

\begin{table*}[t]
\centering
\caption{Stellar Abundances Approaching Ionization Equilibrium with $1\sigma$ Uncertainties \label{table:IEresults}}
\begin{tabular}{l c c c c c}
\tableline\tableline
HIP\,ID & [Fe\,I/H] & [Fe\,II/H] & [O/Fe] & [Mg/Fe] & [Ca/Fe]\\
\tableline
  46120 & $-2.22\pm0.07$ & $-2.18\pm0.23$ & ... & $0.08\pm0.15$ & $0.41\pm0.08$\\
  54639 & $-2.50\pm0.07$ & $-2.46\pm0.19$ & ... & ... & $0.33\pm0.07$\\
  80679 & $-2.50\pm0.07$ & $-2.46\pm0.18$ & ... & ... & $0.38\pm0.07$\\
  87062 & $-1.56\pm0.07$ & $-1.54\pm0.18$ & $0.11\pm0.05$ & $0.47\pm0.25$ & $0.21\pm0.03$\\
  87788 & $-2.66\pm0.07$ & $-2.60\pm0.18$ & ... & ... & $0.79\pm0.05$\\
  98492 & $-1.40\pm0.07$ & $-1.35\pm0.18$ & $0.71\pm0.06$ & $0.41\pm0.15$ & $0.20\pm0.03$\\
103269 & $-1.83\pm0.07$ & $-1.84\pm0.18$ & $0.69\pm0.08$ & $0.14\pm0.15$ & $0.22\pm0.04$\\
106924 & $-2.23\pm0.07$ & $-2.25\pm0.18$ & ... & $0.27\pm0.15$ & $0.40\pm0.04$\\
108200 & $-1.83\pm0.07$ & $-1.79\pm0.18$ & ... & $0.05\pm0.15$ & $0.14\pm0.04$\\
\tableline
\end{tabular}
\vspace{0.25cm}
\end{table*}

\subsection{$\langle$3D$\rangle$ Model Atmospheres \label{sec:3Dmodels}}
Although 1D stellar atmosphere models have improved over the years, for example by including updated atomic and molecular line list data, opacities, and a variety of chemical compositions \citep[e.g.][]{Kurucz,Gust2008}, there are still issues with their assumption of full radiative equilibrium for dwarf metal-poor stars \citep{Magic2013} and their dependence on an adjustable microturbulent velocity parameter inherent in the underlying mixing-length 
theory.  Here, we use the Stagger grid of $\langle$3D$\rangle$ hydrodynamic model atmospheres \citep{Magic2013} to explore the effects of 3D model atmospheres on our derived abundances.  We remind the reader that the $\langle$3D$\rangle$ models are horizontally averaged versions of the full-3D hydrodynamic models, as described by \citep{Magic2013}. 

For the Sun, we employed the Fe~I lines used in the full 3D analysis of \citet{Asplund2000} with the EWs measured here from the Kurucz solar atlas, and the \citet{Magic2013} $\langle$3D$\rangle$ solar model.  This analysis gave a solar Fe~I abundance of 7.48 dex, higher than the 3D \citet{Asplund2000} value of 7.44 dex.  Interestingly, the Kurucz solar model gives an Fe abundance of 7.44 dex from the same lines, in agreement with the full 3D results.

\begin{figure*}[t]
\centering
\includegraphics[width=0.75\textwidth]{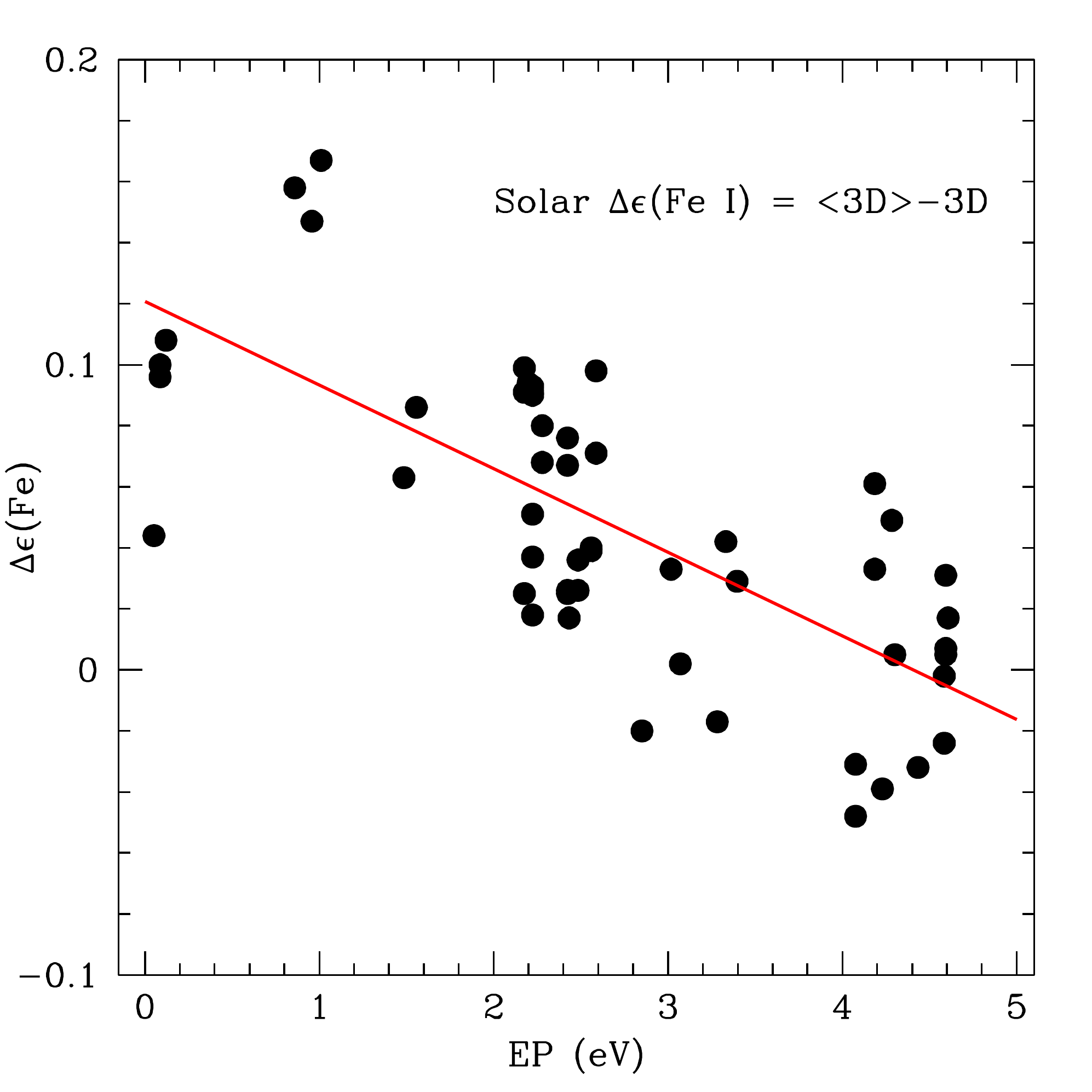}
\caption{A plot showing the difference between Fe~I abundances computed here using the solar $\langle$3D$\rangle$ model from \citet{Magic2013} and the full 3D hydrodynamic model of \citet{Asplund2000}.  The difference slope is independent of log\,gf value of the lines, and suggests that the $\langle$3D$\rangle$ model is 150K hotter than the full 3D model.  
\label{fig-3d3dsolarFeI}}
\vspace{0.25cm}
\end{figure*}

The line-by-line ratio of the individual \citet{Asplund2000} Fe~I 3D abundances to those computed with the \citet{Magic2013} $\langle$3D$\rangle$ solar model showed a strong slope with line EP, as seen in Figure~\ref{fig-3d3dsolarFeI}.  This suggests that the $\langle$3D$\rangle$ model is $\sim$150\,K hotter than the full 3D model.  Note that the slope in Figure~\ref{fig-3d3dsolarFeI} cannot be due to errors in log\,gf values, as these cancel-out in the $\langle$3D$\rangle$ to 3D ratio. Unfortunately, the slope with EP means that the derived Fe~I abundance, relative to the full 3D model, depends on the EP of the line employed and adds significant uncertainty to abundance measurements.

If we employ the $\langle$3D$\rangle$ excitation equilibrium to determine the model T$_{\rm eff}$ for our program stars we must consider that the solar $\langle$3D$\rangle$ excitation temperature is cooler than the actual solar T$_{\rm eff}$ by 150\,K.  Such a cool solar $\langle$3D$\rangle$ model results in lower Fe~I abundances, by 0.08\,dex, and an absolute mean Fe I abundance of 7.36\,dex, with an rms scatter about the mean of 0.06 dex, for the lines used in the \citet{Asplund2000} analysis.

Clearly, the horizontally averaged $\langle$3D$\rangle$ solar model atmosphere \citep{Magic2013} fails to reproduce the full 3D Fe~I abundances results of \citet{Asplund2000}.  We note that this is cannot be due to differences in saturation between the $\langle$3D$\rangle$ and 3D models, because the trend in Figure~\ref{fig-3d3dsolarFeI} is unchanged when only weak lines are considered and there is no clear trend in a plot of abundance differences with EW.

In order to estimate the effect of the $\langle$3D$\rangle$ models on our results, we have computed line-by-line differential spectroscopic T$_{\rm ex}$, Fe~I and Fe~II abundances using the $\langle$3D$\rangle$ models for our standard star, HIP\,66815, and one representative program star, HIP\,103269, which has good S/N and is slightly more metal-rich than many of the other program stars, so its lines are stronger, making the comparison of the effects of Kurucz versus $\langle 3D\rangle$ models easier to identify.  A comparison of these $\langle 3D\rangle$ results with those computed with Kurucz models is presented in Table~\ref{table:3DKurFeH}.  Presently, the uncertainties in Table~\ref{table:3DKurFeH} include random errors only and so should be combined, in quadrature, with the systematic errors discussed in section~\ref{sec:Errors}.

For our standard star, HIP\,66815, the line-by-line differential abundance, relative to the Sun, using the \citet{Magic2013} $\langle$3D$\rangle$ models, gave Fe~I line excitation equilibrium for a model with T$_{\rm eff}$=5970\,K; this is 120\,K hotter than the excitation temperature derived using the Kurucz models.  Notably, this higher temperature agrees well with the photometric V$-$K temperature derived using the \citet{Cas2010} calibration.  

Our $\langle$3D$\rangle$ abundance analysis of HIP\,66815 gave [Fe/H]=$-$0.66 dex from 84 Fe~I lines, but $\mathrm{[Fe/H]}=-0.59$\,dex from 13 Fe~II lines.  The scatter of both Fe~I and Fe~II lines, at 1$\sigma$=0.060 dex, indicates an error on the
mean of 0.017 and 0.007 dex, respectively.  Thus, the $\langle$3D$\rangle$ model for HIP~66815, with the appropriate photometric gravity, did not provide ionization equilibrium.  Compared to the HIP\,66815 results from the Kurucz models, the $\langle$3D$\rangle$ models gave T$_{\rm eff}$ hotter by 120\,K, [Fe~I/H] higher by 0.06 dex , and [Fe~II/H] higher by 0.03 dex.  Neither grid achieved ionization equilibrium for the isochrone gravities, although the $\langle$3D$\rangle$ models gave slightly closer Fe~I and Fe~II abundances.

For our program star, HIP\,103269, the $\langle$3D$\rangle$ model suggested an excitation temperature of 5580\,K compared to 5300\,K for the Kurucz model.  Thus, the difference in excitation temperature between the two grids of model atmospheres appears to increase toward lower metallicity.  For both HIP\,66815 and HIP\,103269 the Kurucz model excitation temperatures are close to the RM05 photometric values, while the $\langle$3D$\rangle$ model excitation temperatures are close to the \citet{Cas2010} scale.

We found the Fe abundance of HIP\,103269 using the $\langle$3D$\rangle$ model of [Fe~I/H]=$-$1.77 dex and $\mathrm{[Fe~II/H]}=-1.64$\,dex.  Thus, we again find that neither model grid obtains Fe ionization equilibrium using the gravities indicated by the DSEP isochrones, although the situation is slightly better for the $\langle$3D$\rangle$ models.  Furthermore, the $\langle$3D$\rangle$ [Fe~I/H] scale is higher than that of the Kurucz grid by $\sim$0.07 dex.

\begin{table*}[t]
\centering
\caption{Comparison of $\langle$3D$\rangle$ and Kurucz [Fe/H] Results \label{table:3DKurFeH}}
\begin{tabular}{r c c c |c c c }
\tableline\tableline
HIP\,ID & T$_{\rm ex}$ (Kurucz) & [Fe\,I/H] & [Fe\,II/H] & T$_{\rm ex}$ ($\langle$3D$\rangle$)   & [Fe\,I/H] & [Fe\,II/H] \\  
\tableline
66815 &  5850 & $-$0.723$\pm$0.006 & $-$0.587$\pm$0.010 & 5970 & $-$0.66$\pm$0.007 & $-$0.59$\pm$0.017\\
103269 &  5300 & $-$1.85$\pm$0.009 & $-$1.64$\pm$0.052 & 5580 & $-$1.77$\pm$0.007 & $-$1.64$\pm$0.087 \\
\tableline
\end{tabular}
\end{table*}

While the $\langle$3D$\rangle$ models of \citet{Magic2013} provide abundances slightly closer to ionization equilibrium, the range of T$_{\rm eff}$, log\,g and [M/H] are not sufficient to properly include the parameters of our program stars; in particular, extra models with log\,g=5.00 at low metallicities would be most helpful.

Our calculations for Table~\ref{table:3DKurFeH} show that LTE abundance analysis using $\langle$3D$\rangle$ model atmospheres gives hotter spectroscopic excitation temperatures than the Kurucz models, and that this T$_{\rm ex}$ difference increases towards progressively more metal-poor stars.  This is consistent with the results of \citet{Berg2012}, who concluded that LTE analysis of metal-poor stars using the $\langle$3D$\rangle$ models leads to over-estimated T$_{\rm ex}$ values.  This resulted from large non-LTE corrections, increasing with decreasing line EP, in the cool regions of the upper atmospheres of the low-metallicity $\langle$3D$\rangle$ models.  Conversely, \citet{Berg2012} found small non-LTE corrections, roughly constant with line EP, when using the 1D MARCS models of \citet{Gust2008}.  Based on the work of \citet{Berg2012}, we conclude that the $\langle$3D$\rangle$ models should only be used with non-LTE abundance analyses.  For LTE abundance analysis the 1D results are preferred over $\langle$3D$\rangle$-LTE results; however, this does not mean that 1D LTE results are any closer to the truth than $\langle$3D$\rangle$ models with full non-LTE analysis.

\subsection{Comparison with Literature Abundance Values}
Several previous studies have attempted to determine the abundances of our target stars in their analyses of the chemical composition of metal-poor dwarf stars in the Milky Way.  In performing this literature comparison, which we present in Table~\ref{table:LitComp}, we specifically used the atmosphere parameters and abundances found when we attempted to achieve ionization equilibrium.

\begin{table*}[t]
\centering
\caption{Literature Comparison \label{table:LitComp}}
\begin{tabular}{lcrccc}
\tableline\tableline
Comparison & Group & Star & $\Delta\mathrm{T}_{\mathrm{eff}}$ (K) & $\Delta$log\,g (dex) & $\Delta$[Fe/H] dex\\
\tableline
\citet{RD1998} & 2 & 46120 & -200 & -0.3 & -0.13\\
\citet{Fulb2000} & 1 & 103269 & 25 & -0.4 & -0.13\\
\citet{Grat2003} & 2 &103269 & -85 & -0.38 & -0.08\\
&& 106924 & -241 & -0.43 & -0.22\\
\citet{CP2005} & 2 & 87062 & 400 & -0.3 & 0.11\\
&& 103269 & 25 & -0.4 & -0.13\\
\citet{VF2005} & 1 & 87062 & 245 & 0.0 & -0.09\\
&& 103269 & 200 & -0.09 & -0.06\\
&& 106924 & 83 & -0.19 & -0.28\\
\citet{ICA2010} & 1 & 87788 & 25 & 0.02 & 0.11\\
&& 108200 & -151 & -0.4 & -0.21\\
\citet{Boes2011} & 1 & 87062 & -392 & -0.19 & -0.28\\
\citet{PM2011} & 1 & 103269 & 200 & -0.09 & -0.11\\ 
\citet{Sou2011} & 1 & 87062 & -216 & -0.61 & -0.17\\
\citet{ICA2012} & 2 & 87788 & -599 & -0.7 & -0.49\\
&& 103269 & -562 & -0.8 & -0.34\\
&& 106924 & -566 & -0.8 & -0.41\\
&& 108200 & -204 & -0.4 & -0.13\\
\citet{Roed2014} & 1 & 80679 & -30 & 0.0 & 0.12 ([Fe\,II/H])\\
&& 108200 & 85 & 0.30 & -0.24 ([Fe\,II/H])\\
\tableline
\end{tabular}
\vspace{0.25cm}
\end{table*}

We find a median difference of -0.13 dex between the abundances determined in this study and those from previous studies \citep{RD1998,Fulb2000,Grat2003,CP2005,VF2005,ICA2010,Boes2011,PM2011,Sou2011,ICA2012,Roed2014}.  It is important to note that the direct comparisons made in Table~\ref{table:LitComp} are complicated by the fact that the studies do not use the same methodology nor do they make use of a consistent line list.  As discussed in \citet{Hink2016}, group-to-group discrepancies could be minimized by using a homogeneous method, line list, and atmospheric parameters when performing an abundance analysis.  As such, we have separated our comparisons based on methodology, specifically, the method in which the effective temperature is determined.  Group 1 indicates studies that used the spectroscopic technique of finding $T_\mathrm{ex}$ by minimizing the trend in Fe\,I abundance with EP while Group 2 indicates studies that used photometric colors to determine $T_\mathrm{eff}$.

Recently, \citet{Roed2014} showed that this choice in methodology for determining stellar temperature creates a significant difference when comparing stellar atmosphere parameters and abundances; where essentially no difference is found between \citet{Roed2014} and other spectroscopic $T_\mathrm{ex}$ studies ($\Delta T_\mathrm{ex} = -28\pm161$ K, $\Delta$ log\,g= -0.24$\pm0.41$ and $\Delta\mathrm{[Fe/H]} = -0.13\pm0.22$ dex from 80 stars), much larger and statistically significant differences are found when comparing to studies that use photometric temperatures ($\Delta T_\mathrm{eff} = -185\pm154$ K, $\Delta$log\,g$ = -0.57\pm0.42$ and $\Delta\mathrm{[Fe/H]} = -0.21\pm0.18$ dex from 110 stars).  We find similar results to \citet{Roed2014} in our comparisons to Group 1 studies which use spectroscopic $T_\mathrm{ex}$ with with a mean $\Delta T_\mathrm{ex} = -2\pm184$ K, $\Delta$ log\,g$-0.12\pm0.26$ and $\Delta\mathrm{[Fe/H]}=-0.14\pm0.11$ dex.  We also find larger differences between our study and those in using photometric temperatures in Group 2 with a mean $\Delta T_\mathrm{eff} = -226\pm306$ K, $\Delta$ log\,g$-0.50\pm0.19$ and $\Delta\mathrm{[Fe/H]}=-0.20\pm0.17$ dex.

\subsection{Comparison to DSEP Isochrones}
There are several important implications of this work that stem from the overall improvement of accuracy in low metallicity stellar models.  The accuracy of these models can be tested using metal-poor MS stars with accurate abundances and parallax measurements.  We briefly discuss the level of agreement between our target star observations and the model isochrones from DSEP, but a more detailed analysis can be found in \citet{Chab2016}.

Eight of the stars used in this study also have accurate \emph{HST} parallaxes provided in \citet{Chab2016} making them suitable candidates for MS-fitting.  We performed a reduced $\chi^2$ analysis in V and I similar to that done in the native HST filters by \citet{Chab2016} in which the location of each star in a CMD is compared to a 12~Gyr isochrone with the given star's metallicity.  The choice of age for the comparison isochrones should not affect the MS-fitting as the location of the MS is independent of age for a given metallicity.  We find the median deviation of the isochrones from the target star observations to be $1.14\sigma$ for isochrones constructed using the \citet{VC2003} color-temperature relation.  The metallicity range of these stars may seem too large to be used in MS-fitting; however, it is important to note that, compared to the alternative of having only a single star from which to base our calibrations, this is a significant improvement.

\begin{figure*}[t]
\centering
\includegraphics[width=0.75\textwidth]{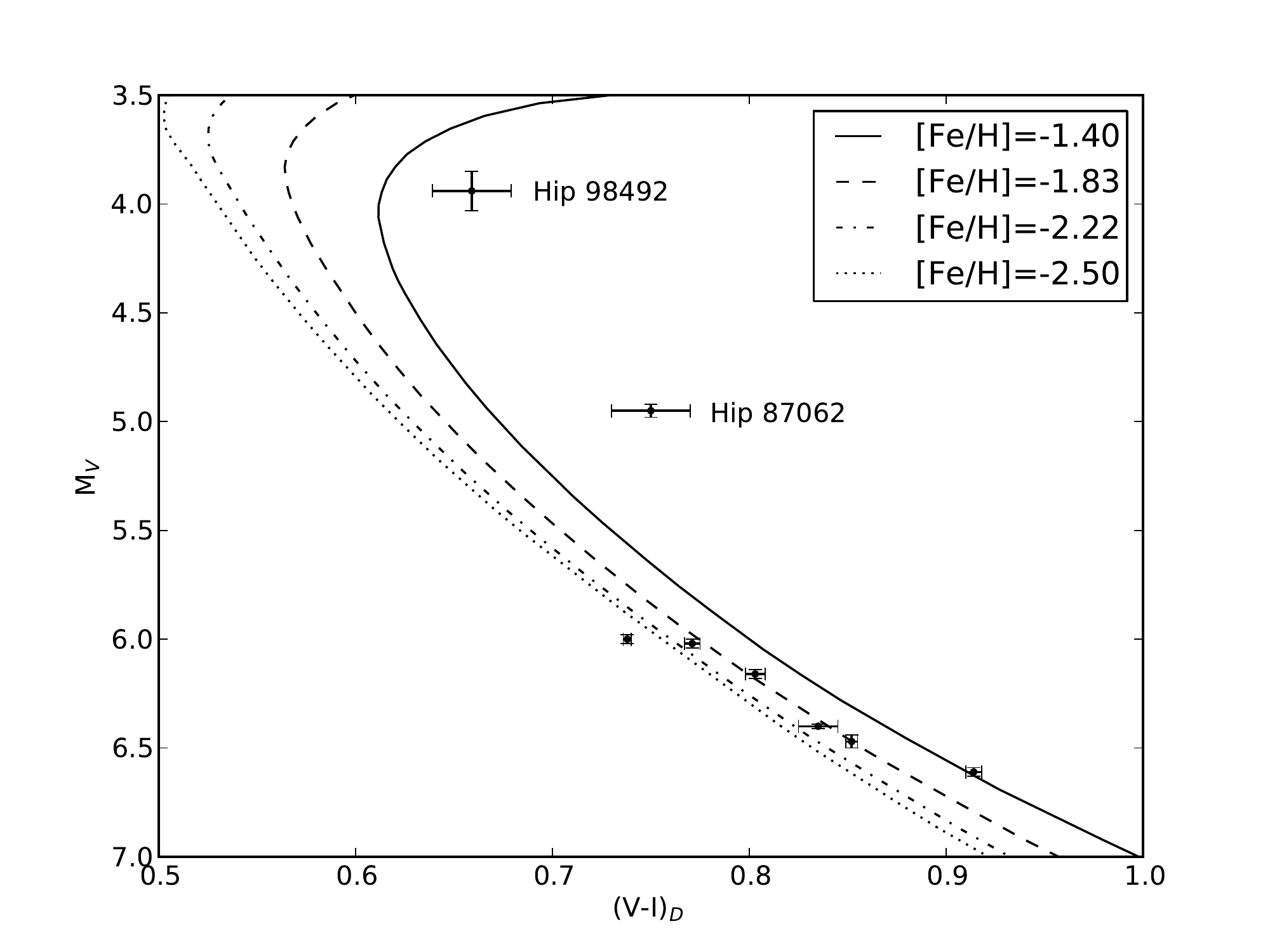}
\caption{The CMD locations (in absolute magnitude and dereddened color) of eight of the target stars are compared to theoretical isochrones in the range of metallicities covered in this study.  There is good agreement for six of the eight stars.  The outliers, HIP\,87062 and HIP\,98492, are both effected by enhanced reddening.  HIP\,87062 may be an unresolved binary while HIP\,98492 is likely a sub-giant branch star. \label{fig:TargetIsochrones}}
\vspace{0.25cm}
\end{figure*}

In Figure~\ref{fig:TargetIsochrones} we show the absolute magnitude and reddening corrected colors of these eight stars in a CMD compared to 12~Gyr DSEP isochrones for their respective metallicities; we exclude HIP\,80679 as it did not have an HST parallax measurement at the time this analysis was performed.  HIP\,98492 and HIP\,87062 have been labeled in the figure as they are obvious outliers.  The low isochrone log\,g we find for HIP\,98492 along with a bright absolute magnitude ($M-V<5.5$ mag) suggests that this star is actually a sub-giant, while the overly red color of HIP\,87062 suggests that either we have underestimated the reddening by at least 0.1 mag or that HIP\,87062 is an unresolved binary that was not filtered out properly.  On the other hand, the agreement between the observed and predicted colors and magnitudes of the remaining six stars in Figure~\ref{fig:TargetIsochrones} suggests that, although we may be uncertain about the underlying atmospheric parameters, the DSEP models provide a good representation of these metal-poor MS stars and can be used with confidence in the MS-fitting of globular clusters.

\section{Discussion \label{sec:Discussion}}
We performed a detailed, line-by-line, differential chemical abundance analysis, ultimately relative to the Sun, for nine metal-poor dwarf stars, near $\mathrm{[Fe/H]}=-2.0$\,dex.  A virtue of this method is the results are independent of line gf values.  Additionally, use of this method should lessen the abundance uncertainty due to effects which may not be included in 1D LTE models (e.g. unidentified blends, non-LTE, 3D and chromospheres).

The spectrum synthesis program used here, MOOG \citep{Sned1973}, includes reliable van der Waals damping constants from \citet{Bark2000}, enabling calculation of synthetic line profiles for damped lines.  However, we found it difficult to distinguish between subtle damped wings and weak blending features for lines larger than $\sim$130 $\mathrm{m\AA}$  in the intermediate comparison star, HIP\,66815, and the Sun and therefore chose not to consider lines with EWs larger than this. 

For a robust abundance analysis we targeted unsaturated lines, on the linear portion of the curve of growth, where the EW is directly proportional to the number of absorbers.  Because of the large metallicity difference between the Sun and the target stars, line-by-line differences for unsaturated lines could only be accomplished using an intermediate standard.  We chose the metal-poor MS star HIP\,66815, with [Fe/H]$\sim$-0.7 dex, as our standard.  However, we note an additional standard, near [Fe/H]$\sim$-1.5 dex, would have been helpful; certainly, there would have been more unsaturated lines with which to compute abundance differences.

For our model atmosphere parameters we chose to consider both spectroscopic and  photometric indicators.  We ultimately chose the model atmosphere T$_{\mathrm{ex}}$ that gave a flat trend of differential abundance with EP, $\Delta\varepsilon$, of Fe~I lines.  We  employed two sets of gravities, the first of which is given by the Dartmouth Stellar Evolution Code, for a 12~Gyr isochrone with [Fe/H]=$-$2, and [$\alpha$/Fe]=+0.4 dex, the second being the surface gravity that gave ionization equilibrium.  

The average non-LTE corrected [Ca/Fe] ratio for our program stars is $+$0.33\,dex, which is close to values seen in halo stars, near $+$0.35 to $+$0.40 dex \citep[e.g.][]{McW97,Cohen04}.  If we remove the highest [Ca/Fe] ratio, [Ca/Fe] = 0.79 dex for HIP\,87788, then we find an average [Ca/Fe] = 0.26 dex.  Since the differential [Ca/Fe] non-LTE correction is estimated to be $+0.12$ dex, then the non-LTE [Ca/Fe] for the 8 stars is $+0.40$ dex, consistent with the MW halo \cite[and references therein]{Sneden2004}.  We conclude that our use of the AODFNEW, $\alpha$-enhanced, Kurucz model atmospheres is reasonably consistent with the spectra.  We note that HIP\,108200 has the lowest non-LTE corrected [Ca/Fe] ratio, at $+$0.14 dex; thus, it is possible that this star belongs to the category of $\alpha$-poor halo stars identified by \citet{NS97,NS10}.  Further chemical composition studies of HIP\,108200 and HIP\,87788 would be useful to check these findings.

We have found in this analysis that the spectroscopic gravities derived for these stars are smaller than those calculated from isochrones using a photometric color-temperature relation by $\sim0.4$ dex.  Because \citet{Chab2016} provides HST parallaxes for eight of these stars we can use these to calculate the gravities as another comparison.  Using these parallaxes along with the apparent magnitudes and reddenings we can find the absolute magnitudes of the stars to be used in the following relation:
\[\mathrm{log }g = 0.4(M_V+BC_V-M_\mathrm{bol,\odot})\mathrm{log}g_\mathrm{\odot}+\]
\[4\mathrm{log}(T_\mathrm{eff}/T_\mathrm{eff,\odot})\mathrm{log}(m/m_\mathrm{\odot})\]
where the bolometric corrections (BC) are interpolated from the Table 3 of \citet{Flower1996} assuming a mass of 0.8\,M$_\mathrm{\odot}$, M$_\mathrm{bol,\odot}=4.74$, $T_\mathrm{eff,\odot} = 5780$\,K, $\mathrm{log}g_\mathrm{\odot}=4.44$\,dex.  In doing so we find the spectroscopic log\,g to be lower than those determined using stellar parallaxes by $0.31\pm0.14$ dex.  This result is similar to that found in \citet{Roed2014}, suggesting that the determination of low log\,g values might simply be an artifact of determining stellar atmosphere parameters spectroscopically, possibly due to systematic error in the 1D model atmospheres.  If one takes the spectroscopic T$_{\mathrm{ex}}$ and log\,g as correct, then the disagreement with the theoretical isochrones would indicate a very large error in the isochrone stellar radii.  However, the comparison between observations and theoretical DSEP isochrones, in color-magnitude space, is reasonably good, with only small color shifts, of order 0.02 mag, required to bring them into agreement, supporting the isochrone results.

When ionization equilibrium is not enforced the Fe~II line abundances are 0.1 to 0.3 dex higher than our Fe~I abundances.  While an increase in T$_{\mathrm{eff}}$ of $\sim$300 K could bring the average of these Fe~I and Fe~II abundances into agreement, and would approximately satisfy photometric color-temperature relations (Alonso et al. 1999; Ramirez \& Gonzalez 2005; Casagrande et al. 2010), such an increase in T$_{\mathrm{eff}}$ would result in a strong slope of abundance with EP and a large dispersion on the mean Fe abundance.  We also note that the error analysis of Table~\ref{table:Errors} shows that a temperature increase of only 150\,K for a typical program star, HIP\,103269, would result in a decrease in the [Ca/Fe] ratio by 0.2 dex, inconsistent with the normal halo [$\alpha$/Fe] ratio.  Thus, the [$\alpha$/Fe] ratios do not favor an increase of model atmosphere T$_{\rm eff}$ to resolve the Fe ionization equilibrium problem.  Because of this, we do not believe that our spectroscopically derived $T_\mathrm{ex}$ are the source of our ionization imbalance nor can the difference be explained by non-LTE over-ionization of Fe\,I because the non-LTE corrections are quite small for our stars, in the range 0.01 to 0.03 dex \citep[e.g.][]{Lind2012}, at least in 1D.

We explored the possibility that 3D hydrodynamical model atmospheres may bring the isochrones and abundance analysis results into better agreement.  Indeed, absolute analysis experiments \citep[e.g.][]{Berg2012} of $\langle$3D$\rangle$ atmospheres \citep[e.g.][]{Magic2013}, give hotter temperatures, based on EP, than the corresponding 1D Kurucz models.  In our analysis of the solar Fe lines using the \citet{Magic2013} $\langle$3D$\rangle$ atmospheres we compared the $\langle$3D$\rangle$ abundances with the full 3D results of \citet{Asplund2000} and found a trend with EP showing the $\langle$3D$\rangle$ excitation temperature is 150\,K cooler than the actual solar T$_{\rm eff}$.  Our $\langle$3D$\rangle$ analysis of HIP\,66815 and HIP\,103269 showed the $\langle$3D$\rangle$ models gave hotter excitation temperatures than the Kurucz models, ranging from 100\,K to 300\,K going from solar metallicity to [Fe/H]$\sim$$-$2.  The [Fe~I/H] metallicity scale of the $\langle$3D$\rangle$ models is 0.07 dex higher than the Kurucz model grid.  Unfortunately, the use of $\langle$3D$\rangle$ models did not lead to Fe ionization equilibrium, although the higher temperatures alleviated part of the problem.  

It is clear that 1D and $\langle$3D$\rangle$ hydrodynamical models combined with separate 1D non-LTE effects do not yet account for the atmospheres of real metal-poor MS stars.  We note, however, that the non-LTE treatment is more appropriately performed with fully 3D atmospheres, not the averaged $\langle$3D$\rangle$ investigated by \citet{Berg2012}.  Such averaged models are really 1D, where ionizing photons must travel from great depths to distort the excitation and ionization of atoms in cooler regions in upper layers.  In the case of fully 3D non-LTE, hot rising bubbles could emit ionizing photons into horizontally adjacent cool regions, over much smaller distances than for the 1D case, possibly enhancing any non-LTE effects.  Such a situation might lead to the ionization and excitation distortions apparent in this work.

The improved accuracy we were hoping to achieve in this work led us to perform a detailed differential abundance analysis on nine metal-poor dwarf stars, but ultimately highlighted issues regarding the log\,g comparison to isochrones.  Given that separate non-LTE and $\langle$3D$\rangle$ calculations do not predict the corrections needed to resolve the log\,g problem it is still possible that fully 3D hydrodynamics plus non-LTE may be at the root of the problem.  However, with [Fe/H] values known to better than 0.1 dex and photometric observations that match theoretical isochrones within a small window of uncertainty we can now look forward to determining improved distances and ages of the oldest MW GCs.  With Gaia projected to provide even more accurate parallaxes in the coming years, it will only be a matter of time before we can not only solve the log\,g problem but have an extensive set of metal-poor stars to use in future calibrations.

\acknowledgments
Support for this work (proposal number GO-11704 \& 12320) was provided by NASA through a grant from the Space Telescope Science Institute which is operated by the Association of Universities for Research in Astronomy, Incorporated, under NASA contract NAS5-26555.  This material is based upon work supported by the National Science Foundation Graduate Research Fellowship under Grant No. DGE-1313911. Any opinion, findings, and conclusions or recommendations expressed in this material are those of the authors(s) and do not necessarily reflect the views of the National Science Foundation.  The authors also wish to recognize and acknowledge the very significant cultural role and reverence that the summit of Mauna Kea has always had within the indigenous Hawaiian community.  We are most fortunate to have the opportunity to conduct observations from this mountain.


\begin{thebibliography}{1}
\expandafter\ifx\csname natexlab\endcsname\relax\def\natexlab#1{#1}\fi

\bibitem[{Alonso et~al.(1999)}]{Alonso1999} Alonso, A., Arribas, S. \& Martinez-Roger, C. 1999, \aap, 140, 261
\bibitem[{Amarsi et al.(2016)}]{Amarsi2016} Amarsi, A.M., Asplund, M., Collet, R., \&
Leenaarts, J. 2016, \mnras, 455, 3735
\bibitem[{Asplund et~al.(2000)}]{Asplund2000} Asplund, M., Nordlund, \AA , Trampedach, R.,
 \& Stein, R.~F. 2000 \aap, 359, 743
\bibitem[{Athay \& Lites(1972)}]{AthayLites72}  Athay, R.~G., \& Lites, B.~W., 1972, \apj, 176, 809
\bibitem[{Barklem et~al.(2000)}]{Bark2000} Barklem, P.~S., Piskunov, N., \& O'Mara, B.~J. 2000, \aap, 142, 467
\bibitem[{Bensby et~al.(2005)}]{Bensby2005} Bensby, T., Feltzing, S., Lundstr\"om, I., \& Ilyin, I. 2005, \aap, 433, 185
\bibitem[{Bergemann et~al.(2012)}]{Berg2012} Bergemann, M., Lind, K., Collet, R., Magic, Z., Asplund, M. 2012 \mnras, 427, 27
\bibitem[{Bergemann \& Norlander(2014)}]{BN2014} Bergemann, M., \& Nordlander, T. 2014, arXiv:1403.3088B
\bibitem[{Beveridge \& Sneden(1994)}]{BS1994} Beveridge, R.~C., \& Sneden, C. 1994, \aj, 108, 285
\bibitem[{Blackwell et~al.(1995a)}]{Black1995a} Blackwell, D.~E., Lynas-Gray, A.~E. \& Smith, G. 1995, \aap, 296, 217
\bibitem[{Blackwell et~al.(1995b)}]{Black1995b} Blackwell, D.~E., Smith, G. \& Lynas-Gray, A.~E. 1995, \aap, 303, 575
\bibitem[{Boesgaard et~al.(2011)}]{Boes2011} Boesgaard, A.~M., Rich, J., Levesque, E.~M., Bowler, B.~P. 2011 \apj, 743, 140 
\bibitem[{Carreta et~al.(2000)}]{Carr2000} Carretta, E., Gratton, R.~G., \& Sneden, C. 2000 \aap, 356, 238
\bibitem[{Casagrande et~al.(2010)}]{Cas2010} Casagrande, L., Ramirez, I., Melendez, J., Bessell, M., Asplund, M. 2010 \aap, 512, A54
\bibitem[{Casagrande et~al.(2011)}]{Cas2011} Casagrande, L., Schr\"onrich, R., Apslund, M., Cassisi, S., Ram\'irez, I., Mel\'endez, J., Bensby, T., \& Feltzing, S. 2011 \aap, 530, A138
\bibitem[{Castelli \& Kurucz(2004)}]{Kurucz} Castelli, F., \& Kurucz, R.L. 2004, unpublished, (arXiv:astro-ph/0405087)
\bibitem[Chaboyer et~al.(1998)]{chaboyer98}Chaboyer, B., Demarque, P, Kernan, P.J. \& Kraus, L.M.\ 1998, \apj, 494 96
\bibitem[{Chaboyer et~al.(2017)}]{Chab2016} Chaboyer, B., McArthur, B.E., O'Malley, E., Benedict, G.F., Feiden, G.A., Harrison, T.E., McWilliam, A., Nelan, E.P., Patterson, R.J., and Sarajedini, A. 2017 \apj, 835, 152
\bibitem[{Charbonnel \& Primas(2005)}]{CP2005} Charbonnel, C., \& Primas, F. 2005 \aap, 442, 961
\bibitem[{Cohen et~al.(2004)}]{Cohen04} Cohen J.G., Christlieb, N., McWilliam, A., Shectman, S.,
Thompson, I. et al. 2004, \apj, 612, 1107
\bibitem[{Collet et~al.(2011)}]{Coll2011} Collet R., Magic Z., Asplund M., 2011, J. Phys. Conf. Ser., 328, 012003
\bibitem[{Conti et~al.(1967)}]{Conti1967} Conti, P.~S., Greenstein, J.~L., Spinrad, H., Wallerstein, G., Vardya, M.~S. 1967, \apj, 148, 105
\bibitem[{Cox(2000)}]{AQ2000} Cox A.N. in``Allen''s Astrophysical Quantities'', 4th ed.  Publisher: New York: AIP Press; Springer, 2000. Edited by Arthur N. Cox. 
\bibitem[{Drawin(1968)}]{Drawin1968} Drawin H.W. 1968, Z. Physik 211, 404
\bibitem[{Drawin(1969)}]{Drawin1969} Drawin H.W. 1969, Z. Physik 225, 483
\bibitem[{Deemin(1961)}]{Deemin1961} Deemin, T.~J. 1961, \mnras, 123, 237
\bibitem[{Dotter et~al.(2008)}]{Dott2008} Dotter, A., Chaboyer, B., Jevremovic, D., Kostov, V., Baron, E., Ferguson, J.~W. 2008 \apjs, 178, 89
\bibitem[{Fabbian et~al.(2009)}]{Fabb2009} Fabbian, D., Asplund, M., Barklem, P.~S., Carlsson, M., Kiselman, D. 2009 \aap, 500, 1221 
\bibitem[{Flower(1996)}]{Flower1996} Flower, P.~J. 1996, \apj, 469, 355
\bibitem[{Fulbright(2000)}]{Fulb2000} Fulbright, J.~P., 2000 \aj, 120, 1841
\bibitem[{Fulbright et~al.(2007)}]{Fulb2007} Fulbright, J.~P., McWilliam, A., \& Rich, R.~M. 2007 \apj, 661, 1152
\bibitem[{Gaia Collaboration(2016)}]{Gaia} Gaia Collaboration 2016 (arXiv:1609.04153)
\bibitem[{Garz et~al.(1969)}]{Garz1969} Garz, T., Holweger, H., Kock, M., \& Richter, J. 1969, \aap, 2, 446
\bibitem[{Gehren et~al.(2006)}]{Gehren2006} Gehren, T., Shi, J.~R., Zhang, H.~W., Zhao, G., \& Korn, A.~J. 2006, \aap, 451, 1065
\bibitem[Gratton  et~al.(1997)]{gratton97} Gratton, R. G., et al.\ 1997, \apj, 491, 749
\bibitem[{Gratton et~al.(2003)}]{Grat2003} Gratton, R.~G., Carretta, E., Claudi, R., Lucatello, S., Barbieri, M. 2003 \aap, 404, 187
\bibitem[Grundahl  et~al.(2002)]{grundahl02}Grundahl, F., Stetson, P. B., \& Andersen, M. I. 2002, \aap, 395, 481
\bibitem[{Gustafsson et~al.(2008)}]{Gust2008}Gustaffson, B., Edvardsson, B., Eriksson, K., Jorgensen, U.~G., Norlund, A., \& Plez, B. 2008, \aap, 486, 951
\bibitem[{Heiter et al.(2015)}]{Heiter2015} Heiter, U., Jofre, P., Gustafsson, B., Korn, A.J., Soubiran, C., \& Thevenin, F. 2015, \aap, 582, A49
\bibitem[{Hinkel et~al.(2016)}]{Hink2016} Hinkel, N.~R., Young, P.~A., Pagano, M.~D., et~al. 2016, \apjs, 226, 4
\bibitem[{Hog et~al.(2000)}]{Hog2000} Hog, E., Fabricius, C., Makarov, V.~V., Urban, S., Corbin, T., Wycoff, G., Bastian, U., Schwekendiek, P., Wicenec, A. 2000, \aap, 355, 27 
\bibitem[{Holweger et~al.(1995)}]{Holweger1995} Holweger, H., Kock, M. \& Bard, A. 1995, \aap, 296, 233
\bibitem[{Huang et~al.(2015)}]{Huang2015} Huang, Y., Liu, X.-W., Yuan, H.-B., Xiang, M.-S., Chen, B.-Q., \& Zhang, H.-W., 2015, \mnras, 454, 2863 
\bibitem[{Ishigaki, Chiba \& Aoki(2010)}]{ICA2010} Ishigaki, M., Chibo, M., \& Aoki, W. 2010 \pasj, 62, 143
\bibitem[{Ishigaki, Chiba \& Aoki(2012)}]{ICA2012} Ishigaki, M., Chibo, M., \& Aoki, W. 2012 \apj, 753, 64
\bibitem[{Koch \& McWilliam(2008)}]{KM2008} Koch, A. \& McWilliam, A. 2008 \aj, 135, 1551
\bibitem[{Koch \& McWilliam(2011)}]{KM2011} Koch, A. \& McWilliam, A. 2011 \aj, 142, 63
\bibitem[{Kopp \& Lean(2011)}]{KL2011} Kopp, G. \& Lean, J.L. 2011 Geophysical Research Letters, 38, L01706
\bibitem[{Latham et al.(2002)}]{Lath2002} Latham, D.~W., Stefanik, R.~P., Torres, G., Davis, R.~J., Mazeh, T., Carney, B.~W., Laird, J.~B., Morse, J.~A. 2002 \aj, 124, 1144
\bibitem[{Lind et~al.(2012)}]{Lind2012} Lind, K., Bergemann, M., \& Asplund, M. 2012, \mnras, 427, 50
\bibitem[{Lindegren et~al.(2016)}]{Lindegren2016} Lindegren, L., et al. 2017, doi:10.1051/0004-6361/201628714; arXiv:1609.04303
\bibitem[{Magic et~al.(2013)}]{Magic2013} Magic, Z., Collet, R., Asplund, M., Trampedach, R., Hayek, W., Chiavassa, A., Stein, R.~F., Nordlund, \AA 2013 \aap, 557, 26
\bibitem[{Mashonkina et~al.(2007)}]{Mash2007} Mashonkina, L., Korn, A.~J., Przybilla, N. 2007, \aap, 461, 261
\bibitem[{Mashonkina (2014)}]{Mash2014} Mashonkina, L. 2014, in ``Setting the scene for Gaia and LAMOST'', Proceedings of the International Astronomical Union, IAU Symposium, Volume 298, pp. 355-365
\bibitem[{McWilliam \& Rich(1994)}]{McWR1994} McWilliam, A. \& Rich, R.~M., 1994 \apjs, 91, 749
\bibitem[{McWilliam et~al.(1995)}]{McW1995} McWilliam, A., Preston, G.~W., Sneden, C., Searle, L. 1995 \aj, 109, 2757
\bibitem[{McWilliam(1997)}]{McW97} McWilliam, A., ARAA, 35, 503
\bibitem[{McWilliam et~al.(2008)}]{McW2008}McWilliam, A., Matteucci, F., Ballero, S., Rich, R.~M., Fulbright, J.~P., Cascutti, G. 2008 \aj, 136, 367
\bibitem[{Mermilliod(1986)}]{Merm1986} Mermilliod, J.C. 1986, Catalogue of Eggen's UBV Data, 0
\bibitem[{Michalik, Lindgren \& Hobbs(2015)}]{MLH2015} Michalik, D., Lindengren, L. \& Hobbs, D. 2015 \aap, 574, 115
\bibitem[{Moore et~al.(1966)}]{NBS61} Moore, C.~E., Minnaert, M. G., \& Houtgast, J. 1966, The Solar Spectrum, 2935~\AA~ to~8770~\AA (NBSMonog. 61, Washington, DC: NBS)
\bibitem[{Munari \& Zwitter(1997)}]{MunZwit1997} Munari, U. \& Zwitter, T. 1997, \aap, 318, 269
\bibitem[{Nissen \& Schuster(1997)}]{NS97} Nissen, P.E., \& Schuster,W.J., 1997 \aap, 326, 751
\bibitem[{Nissen \& Schuster(2010)}]{NS10} Nissen, P.E., \& Schuster,W.J., 2010 \aap, 511, L10
\bibitem[{Perryman \& ESA(1997)}]{hip} Perryman, M.A.C., \& ESA 1997, The HIPPARCOS and TYCHO catalogues. Astrometric and photometric star catalogues derived from the ESA HIPPARCOS Space Astrometry Mission, ESA SP, 1200
\bibitem[{Petigura \& Marcy(2011)}]{PM2011} Petigura, E.~A. \& Marcy, G.~W. 2011, \apj, 735, 41
\bibitem[{Pont  et~al.(1998)}]{pont98}Pont, F., Mayor, M., Turon, C., \& VandenBerg, D. A. 1998, \aap, 329, 87
\bibitem[{Ramirez \& Melendez(2005)}]{RamMel2005} Ramirez, I. \& Melendez, J. 2005, \apj, 626, 465
\bibitem[Reid(1997)]{reid97}Reid, I. N. 1997, \aj, 114, 161
\bibitem[{Roederer et~al.(2014)}]{Roed2014} Roederer, I.~U., Preston, G.~W., Thompson, I.~B., et~al. 2014, \apj, 147, 136
\bibitem[{Ryan \& Deliyannis(1998)}]{RD1998}Ryan, S.~G., \& Deliyannis, C.~P. 1998 \apj, 500, 398
\bibitem[{Sandage(1964)}]{Sandage1964} Sandage, A. 1964, \apj, 139, 442 
\bibitem[{Sitnova et~al.(2013)}]{Sitnova2013} Sitnova, T.M., Mashonkina, L.I., \& Ryabchikova, T.A. 2013, Astron. Lett., 39, 126.
\bibitem[{Sitnova et~al.(2015)}]{Sitnova2015} Sitnova, T., Mashonkina, L., Chen, Y., et al. 2015, \apj, 808, 148
\bibitem[{Sneden(1973)}]{Sned1973} Sneden, C. 1973 \apj, 184, 839
\bibitem[{Sneden(2004)}]{Sneden2004} Sneden, C. 2004, \memsai, 75, 267
\bibitem[{Sobeck et~al.(2006)}]{Sobeck2006} Sobeck, J.~S., Ivans, I.~I., Simmerer, J.~A., Sneden, C., Hoeflich, P., Fulbright, J.~P., \& Kraft, R.~P, 2006 \aj, 131, 2949
\bibitem[{Sousa et~al.(2011)}]{Sou2011} Sousa, S.~G., Santos, N.~C., Israelian, G., Lovis, C., Mayor, M., Silva, P.~B., Udry, S. 2011 \aap, 526, A99
\bibitem[{Struve \& Elvey(1934)}]{SE1934} Struve, O., \& Elvey, C.~T. 1934 \apj, 79, 409
\bibitem[{Wallerstein(1962)}]{Wall1962} Wallerstein, G. 1962 \apjs, 6, 407
\bibitem[{Winkler(1997)}]{Wink1997} Winkler, H. 1997 \mnras, 287, 481
\bibitem[{Valenti \& Fischer(2005)}]{VF2005} Valenti, J.~A., Fischer, D.~A. 2005 \apjs, 159, 141
\bibitem[{VandenBerg \& Clem(2003)}]{VC2003} VandenBerg, D.~A. \& Clem, J.~L. 2003, \apj, 126, 778
\bibitem[{van Leeuwen(2007)}]{Leeu2007} van Leeuwen, F.  2007,  HIPPARCOS, The New Reduction of the Raw Data, Astrophys. Space Sci. Lib. (Springer), 350\bibitem[{Zhao \& Gehren(2000)}]{ZG2000} Zhao, G., \& Gehren, T. 2000, \aap, 362, 1077

\end{thebibliography}
\end{document}